\documentclass[prb,amsmath,twocolumn,showpacs,superscriptaddress]{revtex4}
\usepackage{array}
\usepackage{booktabs}
\usepackage{tabularx}
\usepackage{color}
\usepackage{wrapfig}
\usepackage{graphicx}
\usepackage{psfrag}
\usepackage{multirow}
\usepackage{subfig}	 
\usepackage[english]{babel} 
\usepackage{amsmath,amsthm,amssymb,amsfonts,latexsym} 
\usepackage[latin1]{inputenc} 
\pdfpageattr {/Group << /S /Transparency /I true /CS /DeviceRGB>>}  
\newlength{\picwidth}
\begin{document}
\newcolumntype{Y}{>{\centering\arraybackslash}X}
\newcommand{\head}[1]{\textnormal{\textbf{#1}}}
\newcommand{\normal}[1]{\multicolumn{1}{l}{#1}}
\newcommand{\ra}[1]{\renewcommand{\arraystretch}{#1}}
\newcommand{\beq}{\begin{equation}}
\newcommand{\eeq}{\end{equation}}
\newcommand{\ben}{\begin{eqnarray}}
\newcommand{\een}{\end{eqnarray}}
\newcommand{\bea}{\begin{array}}
\newcommand{\eea}{\end{array}}
\newcommand{\om}{(\omega )}
\newcommand{\bef}{\begin{figure}}
\newcommand{\eef}{\end{figure}}
\newcommand{\leg}[1]{\caption{\protect\rm{\protect\footnotesize{#1}}}}
\newcommand{\ew}[1]{\langle{#1}\rangle}
\newcommand{\be}[1]{\mid\!{#1}\!\mid}
\newcommand{\no}{\nonumber}
\newcommand{\etal}{{\em et~al }}
\newcommand{\geff}{g_{\mbox{\it{\scriptsize{eff}}}}}
\newcommand{\da}[1]{{#1}^\dagger}
\newcommand{\cf}{{\it cf.\/}\ }
\newcommand{\ie}{{\it i.e.\/}\ }

\title{Phosphorene oxides: bandgap engineering of phosphorene by oxidation}
\author{A. Ziletti }

\affiliation{Department of Chemistry, Boston University, 590 Commonwealth Avenue, Boston Massachusetts 02215, USA}
\author{A. Carvalho}
\affiliation{Centre for Advanced 2D Materials and Graphene Research Centre, National University of Singapore, 6 Science Drive 2, 117546, Singapore}

\author{P. E. Trevisanutto}
\affiliation{Centre for Advanced 2D Materials and Graphene Research Centre, National University of Singapore, 6 Science Drive 2, 117546, Singapore}

\author{D. K. Campbell}
\affiliation{Department of Physics, Boston University, 590 Commonwealth Avenue, Boston Massachusetts 02215, USA}

\author{D. F. Coker}
\affiliation{Department of Chemistry, Boston University, 590 Commonwealth Avenue, Boston Massachusetts 02215, USA}

\author{A. H. Castro Neto}
\affiliation{Centre for Advanced 2D Materials and Graphene Research Centre, National University of Singapore, 6 Science Drive 2, 117546, Singapore}
\affiliation{Department of Physics, Boston University, 590 Commonwealth Avenue, Boston Massachusetts 02215, USA}

\begin{abstract}
We show that oxidation of phosphorene can lead to the formation of a new family of planar (2D) and tubular (1D)
oxides and sub-oxides, most of them insulating.  
This confers to black phosphorus a native oxide that can be used as barrier material and protective layer.
Further, the bandgap of phosphorene oxides depends on the oxygen concentration, suggesting that controlled 
oxidation can be used as a means to engineer the bandgap.
For the oxygen saturated composition, P$_2$O$_5$, both the planar and tubular phases
have a large bandgap energy of about 8.5~eV, and are transparent in the near UV. These two forms of phosphorene oxides are predicted to have the same formation enthalpy as o$^\prime$-P$_2$O$_5$, the most stable of the previously known forms of phosphorus pentoxide. 
\end{abstract}
\maketitle


\section{Introduction}

Phosphorene, a single layer of black phosphorus, is a unique two-dimensional semiconductor
with a nearly-direct gap in the visible range.\cite{rodin2014,tran2014} 
It stands out amongst the family of 2D materials for its orthorhombic 
waved structure with superior flexibility and small Young's modulus,\cite{wei2014}
allowing for strain-driven band structure engineering.\cite{rodin2014}
From the point of view of electronic and optoelectronic applications, it features high carrier mobility
and on-off ratio,\cite{koenig2014,xia2014}
as well as a giant photoresponse in the IR and UV.\cite{buscema2014}

In contrast to graphene, phosphorene is prone to oxidation,\cite{ziletti2014}
which usually leads to degradation of the structure and electronic properties.\cite{rosti2014,castellanos-gomez2014}
At present oxidation is avoided by encapsulating with polymers or other two-dimensional materials
eg. graphene or boron nitride.\cite{rosti2014}

However, there are indications that a phosphate layer grown in a controlled way could be used 
as a protective layer or even as a functional material on its own.
Black phosphorus oxidizes on the surface, seemingly leaving the deepest layers intact.
The non-uniform degradation pattern normally observed, showing as dark patches on the optical microscope image, 
seems to require the interaction with water,\cite{favron,stm}
even though energy-loss spectra shows P$_x$O$_y$ still grows in oxygen-only atmosphere.\cite{farnsworth2014}
In this article, we show that phosphorene oxides and suboxides can exist in monolayer form,
suggesting that there is an opportunity for growth of single-layer or few-layer native oxide
at the black phosphorus surface.

The existence of an insulating native oxide is an added advantage for phosphorene.
Since phosphate glasses,\cite{brow2000} (e.g. monolayer P$_2$O$_{5}$) are transparent in the near UV,
a passivating coating will preserve the optical properties of the underlying phosphorene.
Further, since such a coating it is saturated with oxygen, it prevents oxygen molecules from reaching the pristine phosphorene layers beneath.

Bulk P$_2$O$_{5}$ has at least three known planar polymorphs, a molecular solid\cite{jansen1986} (with P$_4$O$_{10}$ structural units),
and two orthorhombic phases,\cite{arbib1996,stachel1995} along with an amorphous phase.\cite{hullinger1976,galeener1979} 
The thermodynamically most stable form (o$^\prime$-P$_2$O$_5$) is a layered structure belonging to the $Pnma$ space group.\cite{stachel1995}
A monolayer phase, different from the one reported here, has also been proposed by a recent theoretical study.\cite{wang}

In this article, we show that suboxides (P$_4$O$_{n}$, $n<10$) also exist in a variety of layered forms,
most of them insulating, with bandgaps that depend on the oxygen concentration.
This offers the possibility of using oxidation as a means to engineer the bandgap of phosphorene. 
The article is divided in three parts: 
first, we describe the structure and energetics of phosphorene oxides; then, we consider their electronic properties and finally their vibrational spectra.


\section{Computational details}
All first-principle calculations are based on the framework of density functional theory (DFT), as implemented in the {\sc Quantum ESPRESSO} package\cite{qe}. 
We use the PBE\cite{pbe}, PBEsol\cite{PBEsol} and HSE06\cite{hse06} (HSE hereafter) approximations for the exchange and correlation energy. The PBEsol functional is preferred over PBE because it cures the systematic tendency of PBE to overestimate equilibrium lattice constants of solids\cite{PBEsol}. The HSE functional, with its fraction of screened short-ranged Hartree-Fock exchange, yields reasonably accurate predictions for energy bandgaps in semiconductors\cite{heyd2005,janesko2009,henderson2011}. Unless otherwise stated, all quantities reported (e.g. energies, bond lengths and bond angles) are obtained with the PBEsol functional. The electron-ion interaction is described using the projector-augmented wave (PAW)\cite{paw} approach for PBEsol calculations, while norm-conserving Troullier-Martins pseudopotentials\cite{tm-pseudo} are employed in PBE and HSE calculations.  
We employ a plane wave basis set with kinetic energy cutoffs of 70~Ry (280~Ry) to describe the electronic wave functions (charge density).
Vibrational properties including Raman and infrared spectra are calculated with density functional perturbation theory\cite{baroni2001} (DFPT) and the PBEsol functional.
The Brillouin zone is sampled using a $\Gamma$-centered 10$\times$8$\times$1 Monkhorst-Pack (MP) grid\cite{mpgrid} for all but the vibrational calculations, for which a finer 15$\times$12$\times$1 grid is employed. A supercell of 16~\AA\ in the direction perpendicular to the monolayer is used to avoid spurious interactions between periodic replicas. For each oxygen concentration, tens of different configurations are used as starting points for optimization. Then, lattice geometries and atomic positions are relaxed till the forces on each atom are less than 10$^{-3}$~eV/\AA, and the pressure less then 1~kbar, except for the HSE calculations, where we use the PBE lattice parameters.
The tetrahedron smearing method\cite{smear-dos} is used with a Gaussian broadening of 0.04~eV and a 12$\times$9$\times1$ MP grid in the density of states (DOS) calculations. 
However, local and semilocal DFT functionals (such as PBE and PBEsol) are known to underestimate the bandgap because of their missing derivative discontinuity in the exchange-correlation energy across the gap\cite{perdew1982,perdew1983}. The inclusion of a fraction of exact nonlocal short-ranged Hartree-Fock exchange, as done in HSE, partially cures this shortcoming, giving bandgaps in better agreement with experiment\cite{hse06}. 
Thus, we use HSE to calculate the bandgap energy\cite{henderson2011}. 

To further validate the HSE results, we calculate the bandgap of pristine phosphorene and the two phosphorene oxides ($p$-P$_4$O$_{10}$ and  $t$-P$_4$O$_{10}$) using the GW approximation to the electron self-energy within the generalized plasmon pole model.\cite{hedin1965,hybertsen1986} The GW method is state-of-the-art for evaluating quasiparticle bandgaps since it gives values in very good agreement with experiment (typically within 0.2~eV) for a large variety of systems and a broad range of bandgaps\cite{louie1998}. 
The GW calculations are performed in two stages. First, relaxed lattice geometries, equilibrium atomic positions and the electronic ground state are obtained from a DFT calculation with the PBE functional and Troullier-Martins pseudopotentials\cite{tm-pseudo}. Then, the DFT Kohn-Sham orbitals and energies are used to construct both the electronic Green's function, G, and the dynamically screened interaction, W, to evaluate the quasiparticle self-energy $\Sigma \approx i$GW. This method is commonly referred as G$_0$W$_0$. 
To achieve convergence in our GW calculations we follow the procedure proposed by Malone and Cohen.\cite{malone2013} 
For pristine phosphorene, we use 370 bands to evaluate the dielectric matrix $\epsilon$ and the self-energy $\Sigma$, employing an energy cutoff $\epsilon_{\rm cutoff}$ of 12~Ry for the dielectric matrix and a 24$\times$20$\times$1 MP-grid; convergence was however checked with up to 1024 bands and using $\epsilon_{\rm cutoff}$ up to 20~Ry.
For the two phosphorene oxides ($p$-P$_4$O$_{10}$ and  $t$-P$_4$O$_{10}$) we use an energy cutoff $\epsilon_{\rm cutoff}$ of 8~Ry for the dielectric matrix, and 800 and 896 bands to evaluate $\epsilon$ and $\Sigma$, respectively. A 7$\times$4$\times$1 grid for $p$-P$_4$O$_{10}$ and a 4$\times$6$\times$1 grid for $t$-P$_4$O$_{10}$ are employed. Convergence was checked including up to 1792 bands and increasing $\epsilon_{\rm cutoff}$ up to 20~Ry. 
For all the GW calculations we use a supercell of 20~\AA\ in the direction perpendicular to the monolayers and a slab-truncation of the Coulomb potential.\cite{beigi2006} The GW calculations are performed with the ABINIT code.\cite{abinit2009}
Additional details on the convergence study are given in the \emph{Supplemental Material}. 
With these parameters, we conservatively estimate the errors in our GW bandgaps at 0.05 eV for pristine phosphorene and 0.1~eV for $p$-P$_4$O$_{10}$ and $t$-P$_4$O$_{10}$.


\section{Results and discussion}


\subsection{Structure}
First we describe the new family of phosphorene oxides (POs) found in this study.
These are obtained by adding oxygen atoms to phosphorene, maintaining the rectangular lattice symmetry.
We considered all compounds P$_4$O$_{n}$ with $n$ between 1 and 10.
The maximum oxygen concentration of 250\%, or 10 oxygen atoms per unit cell, 
corresponds to the stoichiometry of phosphorus pentoxide (P$_2$O$_5$ or P$_4$O$_{10}$), 
whose known polymorphs were outlined in the introduction\cite{greenwood1997}. 

\begin{figure}[htb]
\centering
    \includegraphics*[trim=0pt 235pt 0pt 240pt, width=8.6cm]{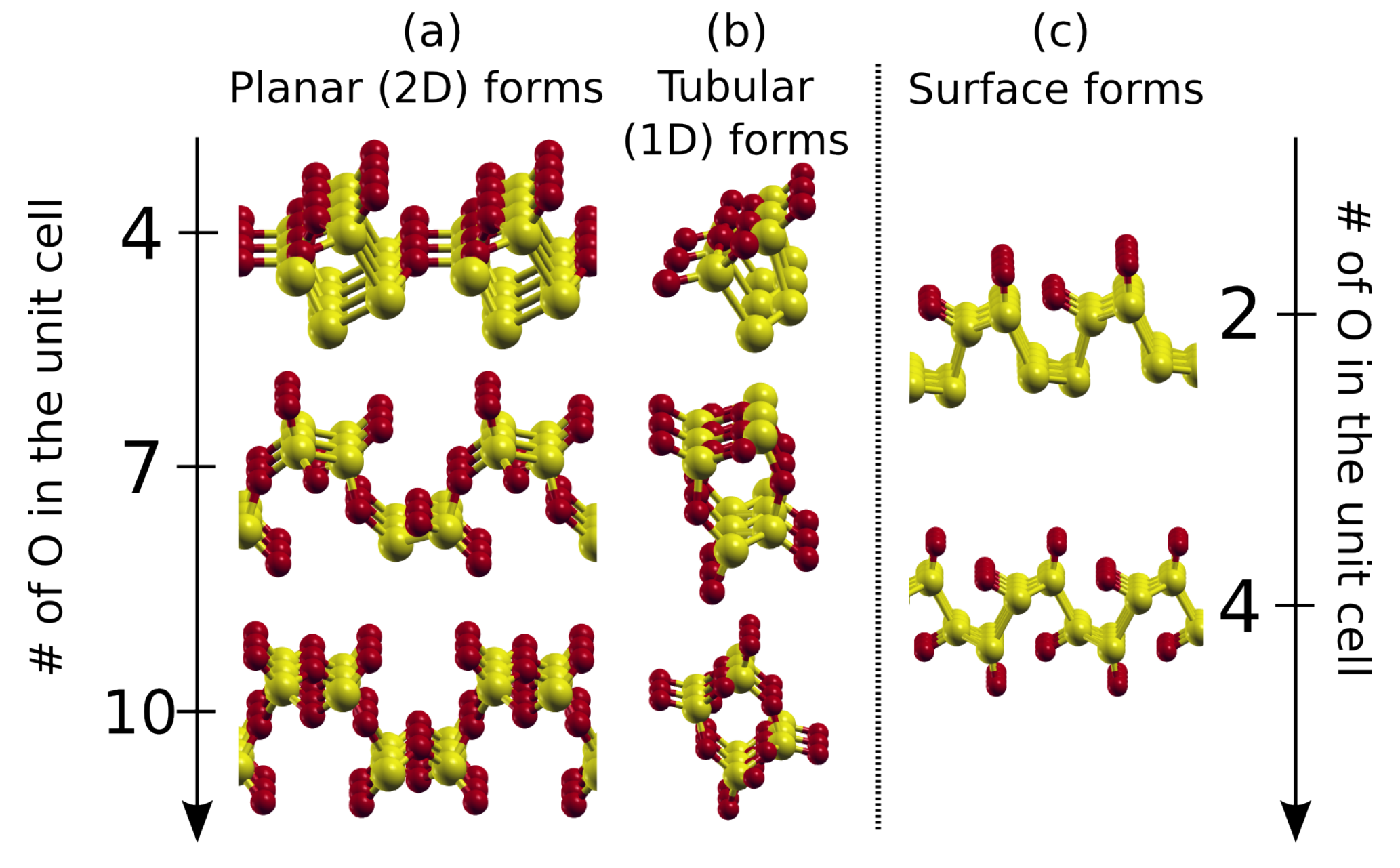}
 \caption{\small Representative structures of phosphorene oxides.}
\label{fig:structures}
\end{figure}

For each oxygen concentration, we use tens of different starting points corresponding to different O atoms arrangements in the pristine phosphorene lattice. After lattice relaxation and geometry optimization, numerous metastable phosphorous oxide (PO) structures are identified. 
We find that, for each oxygen concentration, the most stable POs can be always divided into two very distinctive classes: planar (2D)  and tubular (1D) forms. For convenience, we refer to the planar forms as $p$-P$_4$O$_n$ and to the tubular forms as $t$-P$_4$O$_n$, where $n$ is the number of oxygen atoms in the unit cell. We also identify surface forms $s$-P$_4$O$_n$. Some representative structures are depicted in Fig.\ref{fig:structures}.

\begin{figure}[htb]
\centering
    \includegraphics*[trim=0pt 300pt 0pt 300pt, width=5cm]{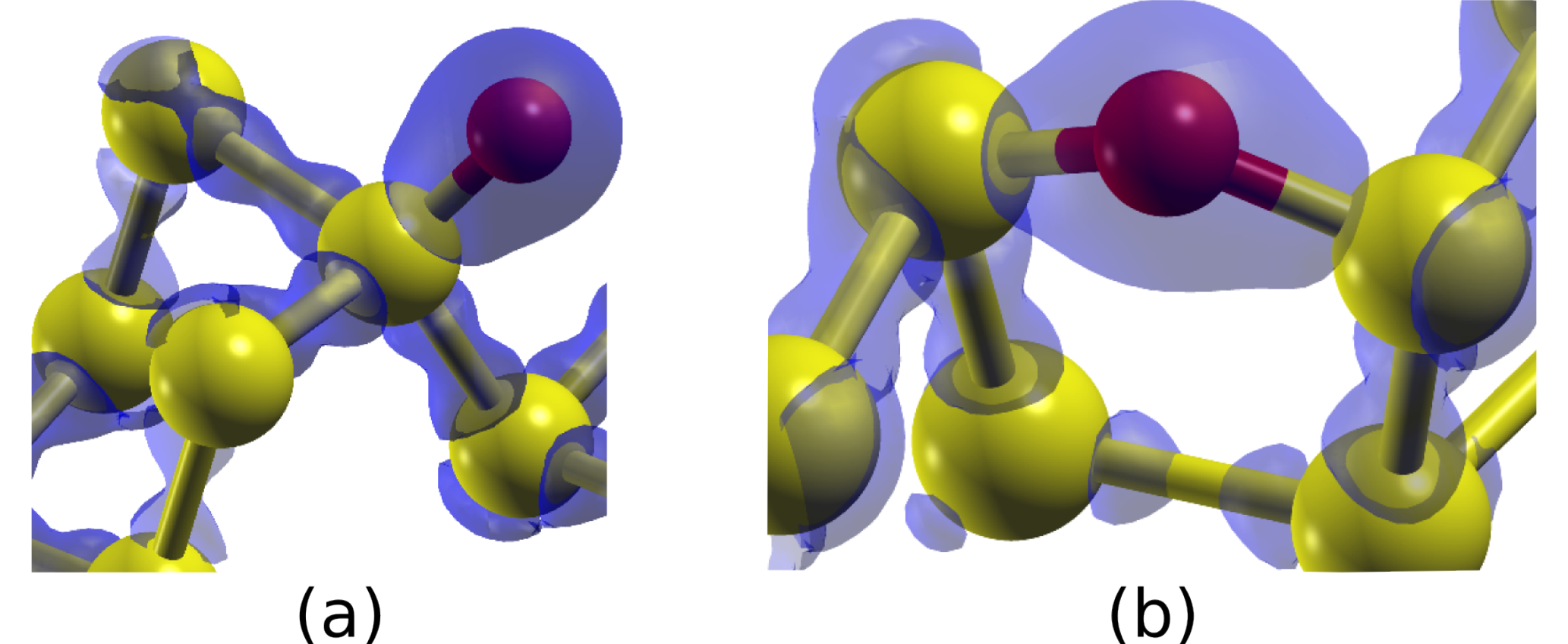}
 \caption{\small Examples of charge density distribution in recurrent motifs in phosphorene oxides: (a) dangling oxygen (b) bridging oxygen. The isosurfaces correspond to $1/10$ of the total charge density.}
\label{fig:motifs}
\end{figure}

Analysis of the local oxygen atom environments that are common to both $p$-P$_4$O$_n$ and $t$-P$_4$O$_n$ forms presented in Fig.\ref{fig:structures} indicates that all the PO structures identified exhibit two P-O motifs: the dangling and bridging oxygen motifs shown in Fig.\ref{fig:motifs}. 

The dangling oxygen motif, represented in Fig.\ref{fig:motifs}a, consists of an O atom that forms a bond with only one P atom. 
The phosphorus lone pair is attracted towards the more electronegative oxygen atom, giving rise to an excess of negative charge localized on the O atom (Fig.\ref{fig:motifs}a). 
The P-O bond is short, strong and polar with bond lengths ranging from 1.44~\AA\ to 1.51~\AA\, depending on phosphorus oxidation number and local environment. \footnote{Oftentimes this bond is classified as a double bond\cite{galeener1978,galeener1979,greenwood1997,brazhkin2011}, although its identification as P=O is not commonly accepted\cite{gilheany1994,rai1994,chesnut1999}. Nevertheless, hereafter we will refer to this bond as P=O for simplicity.}
Similar to what has been found for dangling oxygens on a phosphorene surface\cite{ziletti2014},
in the PO forms identified here, the P=O bonds always point away from the zigzag ridge in which they are chemisorbed, minimizing Coulomb repulsion between the oxygen $p$ orbitals.

The P=O bond length decreases monotonically as the number of oxygens linked to the phosphorus involved in the bond increases. 
Let P$^{(n)}$ be a phosphorus atom that is linked to $n$ oxygens, where $n=1,\dots, 4$. 
With this notation P$^{(1)}$=O defines a double bond in which a phosphorus is linked to only one oxygen, as shown for example in Fig.\ref{fig:motifs}a. 
The average P=O bond lengths for the sequence of P$^{(n)}$ configurations
are given in Table~\ref{table:blength} for the planar and tubular forms.
The longest P=O is found for $n=1$ (where the P atom establishes a P=O bond and three P-P bonds); 
conversely, the shortest P=O are found in $p$-P$_4$O$_{10}$ and $t$-P$_4$O$_{10}$ ($n=4$), 
where the phosphorus atom is bonded to four oxygens.
This monotonic decrease in the P=O bond length with increasing oxygen connectivity has an analogue in the molecular series P$_4$O$_m$ ($m$=6$-$10)\cite{jansen1981,walker1979,clade1994}. 

\begin{table}[htb]
\caption{Average bond lengths for various O linkages in planar and tubular forms (\AA).}
\centering
\begin{tabular}{l c c}
\hline\hline
&planar  &  tubular \\
\hline
P$^{(1)}$=O& 1.508& 1.477\\
P$^{(2)}$=O& 1.476& 1.466\\
P$^{(3)}$=O& 1.459& 1.458\\
P$^{(4)}$=O& 1.445& 1.447\\
\hline\hline
\end{tabular}
\label{table:blength}
\end{table}

In the bridge motif, as the name indicates, the oxygen atom bridges two P atoms, occupying a position close to a P-P bond center (Fig.\ref{fig:motifs}b). 
The electron density is mostly distributed on the two P-O bonds, and only a moderate electron density accumulation is observed on the oxygen. 
Therefore, we expect a much stronger Coulomb repulsion between dangling oxygens than bridging oxygens, especially for high O concentrations. 
The P-O bondlengths in this case are between 1.61~\AA\ and 1.78~\AA, this wide range due to different strain interactions and lattice distortions depending on the oxygen concentration.

\begin{figure}[htb]
\centering
    \includegraphics*[trim=0pt 20pt 0pt 10pt, width=8.6cm]{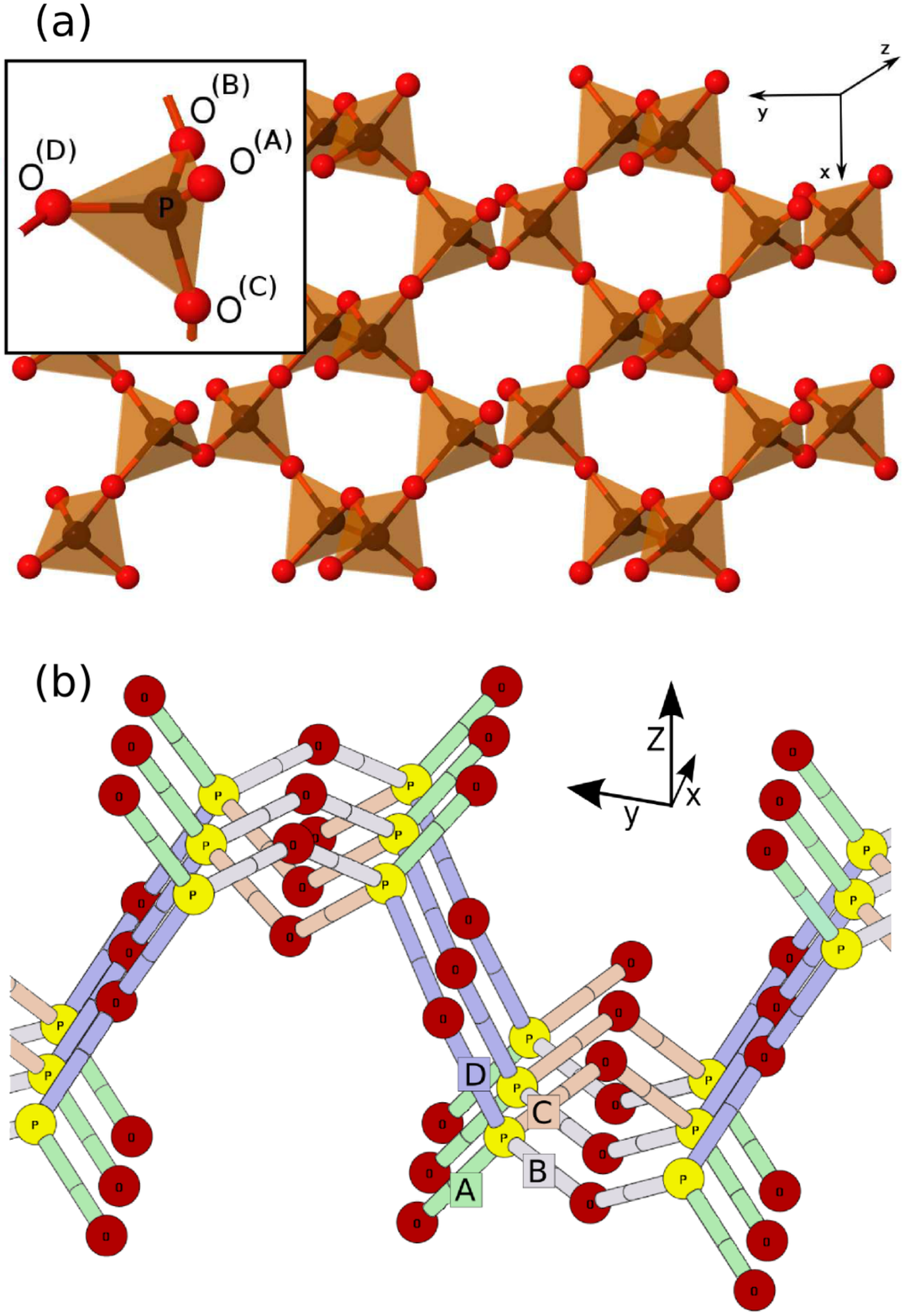}
 \caption{\small Structure of $p$-P$_4$O$_{10}$ (a) Top view. The quasi-tetrahedral PO$_4$ unit is shown in the inset.
(b) Profile view. The four colors indicate four different bondlengths (A, B, C, D). Bonds with the same colour are identical.}
\label{fig:p-p4o10}
\end{figure}

Bridging oxygens can either bond with P atoms from the same or from different zigzag ridges, therefore forming intra-ridge ({\em e.g.} B or C in Fig. \ref{fig:p-p4o10}b) or inter-ridge bridge ({\em e.g.} D in Fig. \ref{fig:p-p4o10}b) structures. 
In particular, inter-ridge bridges are essential for the formation of the planar forms because they effectively link different ridges that would otherwise separate and form 1D tubular chains. 
The formation of oxygen bridges increases considerably the lattice parameters, resulting in deformations as large as 90\% relative to pristine phosphorene for the maximally oxidized forms (see Table \ref{table:latt-param}).

\begin{table}[htb]
\caption{Lattice parameters and deformations relative to pristine phosphorene for representative POs.} 
\centering 
\begin{tabular}{l l  c c  | 
c  c c} 
\hline\hline 

System & XC  & $a$ (\AA) & $b$ (\AA) &  $\Delta$a(\%) & $\Delta$b(\%) \\
  
\hline

Phosphorene  & PBE      & 3.30 & 4.62 & --  & --     \\  
		      & PBEsol & 3.28 & 4.45 & --  & --   \\  
\hline
$p$-P$_4$O$_{10}$ & PBE      & 4.39 & 6.52 & 33 &  41   \\  
				& PBEsol & 4.41 & 6.75 &  34  &  51   \\  
\hline
$t$-P$_4$O$_{10}$ & PBE      & 6.12 & 4.53 &  85  & -1.9    \\  
			       & PBEsol  & 6.28 & 4.58 &  91  &  2.9   \\  
\hline			    
$s$-P$_4$O$_{2}$ & PBE & 3.44 & 4.52 & 4.2  & -2.2    \\  
& PBEsol & 3.41 & 4.37 &  3.9 & -1.8    \\ 
$s$-P$_4$O$_{4}$ & PBE & 3.60 & 4.76 & 9.1  & 3.0    \\  
& PBEsol & 3.57 & 4.85 & 8.2  & 9.0    \\  

\hline\hline 
\end{tabular}
\label{table:latt-param}
\end{table}

\begin{figure}[htb]
\centering
    \includegraphics*[trim=38pt 190pt 0pt 180pt, width=8.6cm]{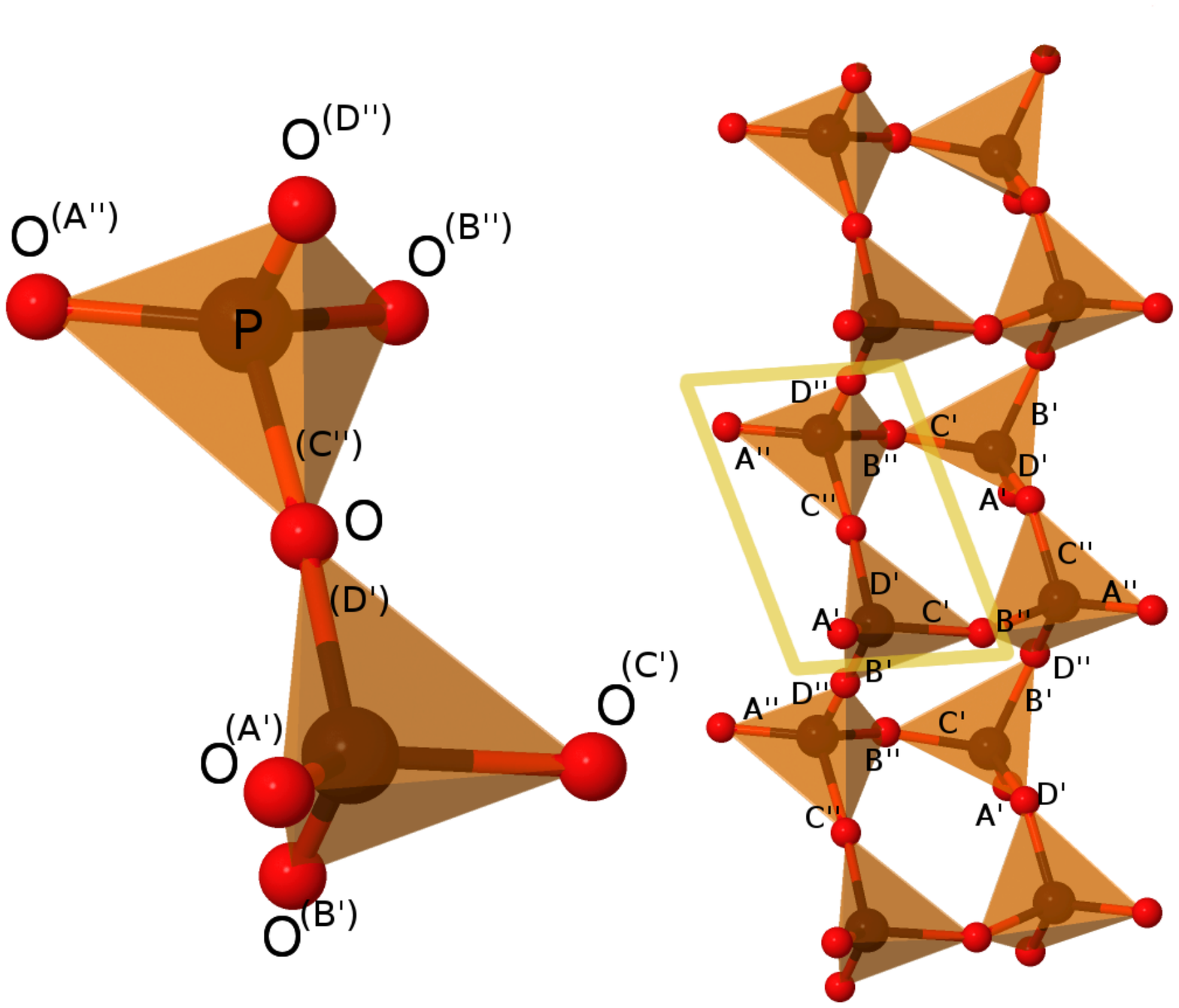}
 \caption{\small Single tubular chain of $t$-P$_4$O$_{10}$. The two inequivalent quasi-tetrahedral PO$_4$ units forming this tubular PO are shown on the left side.}
\label{fig:t-p4o10}
\end{figure}

Dangling and bridging oxygens can occupy numerous positions in the lattice, giving rise to a manifold of metastable, nearly degenerate structures. 
In the following we examine in detail the most stable POs obtained for maximum oxygen concentration: the planar $p$-P$_4$O$_{10}$ and the tubular $t$-P$_4$O$_{10}$ forms. These are shown in Fig.\ref{fig:p-p4o10} and Fig.\ref{fig:t-p4o10}, respectively.

The planar $p$-P$_4$O$_{10}$ is a waved structure, like the parent phosphorene.
It can be built by placing an oxygen atom near each bond-center and near each lone pair.
The space group is $Cm$ ($C_s^3$), and the unit cell contains 4 phosphorus and 10 oxygen atoms. There are four inequivalent P-O bonds, three of which connect the phosphorus atom with bridging oxygens (B,C,D in Fig.\ref{fig:p-p4o10}b), while the forth links the phosphorus to a dangling oxygen (A in Fig.\ref{fig:p-p4o10}b). The respective bond lengths are given in Table~\ref{table:bonds-angles}.
To minimize Coulomb repulsion, the bridging oxygens between P atoms in the same zigzag ridge alternate up and down buckling, forming two rows parallel to the $x$-direction and separated by 1.23~\AA\ in the $z$-direction (see brown and grey bonds in Fig.\ref{fig:p-p4o10}b), one row being above and the other below the $x$-$y$ plane passing through the two P atoms in the zigzag ridge.

Alternatively $p$-P$_4$O$_{10}$ can be seen as consisting of identical quasi-tetrahedral PO$_4$ units,
forming a network of vertex-sharing tetrahedra (Fig.\ref{fig:p-p4o10}a).
The bond lengths of the PO$_4$ tetrahedra are listed in Table \ref{table:bonds-angles}.
The deviations from the ideal tetrahedral bond angles (109.5$^{\circ}$) can be straightforwardly explained in the framework of valence-shell electron pair repulsion (VSEPR) theory\cite{gillespie1957,gillespie1963,gillespie1992}.
The P-O$^{({\rm A})}$ bond has in fact a double bond character, hence the repulsion between electron-pair domains in VSEPR is larger\cite{gillespie1992}, causing larger bond angles relative to the idealized tetrahedral geometry. 

\begin{table}[htb]
\caption{Bondlengths and bond angles for $p$-P$_4$O$_{10}$ and $t$-P$_4$O$_{10}$ calculated with the PBEsol functional.
For the surface oxide forms, O-P$^{({\alpha})}$-P$^{({\beta})}$
is the angle between the  O-P$^{({\alpha})}$ and the P$^{({\alpha})}$-P$^{({\beta})}$ bonds, where P$^{({\alpha})}$ and P$^{({\beta})}$
are nearest neighbors in the same zig-zag ridge.}
\begin{tabularx}{0.46\textwidth}{X X  X X} 
\hline\hline 
\multicolumn{4}{c}{Planar form: $p$-P$_4$O$_{10}$} \\  \hline
\multicolumn{2}{c|}{Bondlengths (\AA)} & \multicolumn{2}{c}{Bond angles ($^{\circ}$)} \\ 
\hline
P=O$^{({\rm A})}$ & \multicolumn{1}{c|}{1.444}  & O$^{({\rm A})}$-P-O$^{({\rm B})}$ &  \multicolumn{1}{c}{115.1}\\
P-O$^{({\rm B})}$& \multicolumn{1}{c|}{1.585} & O$^{({\rm A})}$-P-O$^{({\rm C})}$ & \multicolumn{1}{c}{115.7} \\
P-O$^{({\rm C})}$& \multicolumn{1}{c|}{1.613} & O$^{({\rm A})}$-P-O$^{({\rm D})}$ &  \multicolumn{1}{c}{117.8} \\
P-O$^{({\rm D})}$& \multicolumn{1}{c|}{1.606} &  \\
\hline \hline
\multicolumn{4}{c}{Planar surface form: $s$-P$_4$O$_{2}$} \\  \hline
\multicolumn{2}{c|}{Bondlengths (\AA)} & \multicolumn{2}{c}{Bond angles ($^{\circ}$)} \\ 
\hline
P=O$^{({\rm A})}$ & \multicolumn{1}{c|}{1.477}  & O-P$^{({\rm a})}$-P$^{({\rm b})}$ &  \multicolumn{1}{c}{111.8}\\
P=O$^{({\rm B})}$ & \multicolumn{1}{c|}{1.516}  & O-P$^{({\rm c})}$-P$^{({\rm d})}$ &  \multicolumn{1}{c}{104.1}\\
\hline \hline
\multicolumn{4}{c}{Planar surface form: $s$-P$_4$O$_{4}$} \\  \hline
\multicolumn{2}{c|}{Bondlengths (\AA)} & \multicolumn{2}{c}{Bond angles ($^{\circ}$)} \\
\hline
P=O$^{({\rm A})}$ & \multicolumn{1}{c|}{1.468}  & O-P$^{({\rm a})}$-P$^{({\rm b})}$ &  \multicolumn{1}{c}{111.3}\\
P=O$^{({\rm B})}$ & \multicolumn{1}{c|}{1.510}  & O-P$^{({\rm c})}$-P$^{({\rm d})}$ &  \multicolumn{1}{c}{110.8}\\
\hline \hline
\multicolumn{4}{c}{Tubular form: $t$-P$_4$O$_{10}$}\\ \hline
\multicolumn{2}{c|}{Bondlengths} & \multicolumn{2}{c}{Bond angles} \\ 
\hline
P=O$^{({\rm A^{\prime}})}$& \multicolumn{1}{c|}{1.442} & O$^{({\rm A^{\prime}})}$-P-O$^{({\rm B^{\prime}})}$ & \multicolumn{1}{c}{115.5}\\
P-O$^{({\rm B^{\prime}})}$& \multicolumn{1}{c|}{1.595} & O$^{({\rm A^{\prime}})}$-P-O$^{({\rm C^{\prime}})}$ & \multicolumn{1}{c}{115.6} \\
P-O$^{({\rm C^{\prime}})}$& \multicolumn{1}{c|}{1.599} & O$^{({\rm A^{\prime}})}$-P-O$^{({\rm D^{\prime}})}$ & \multicolumn{1}{c}{118.6} \\
P-O$^{({\rm D^{\prime}})}$& \multicolumn{1}{c|}{1.604} &  \\ \hline
P=O$^{({\rm A^{\prime \prime}})}$& \multicolumn{1}{c|}{1.447} & O$^{({\rm A^{\prime \prime}})}$-P-O$^{({\rm B^{\prime \prime}})}$ & \multicolumn{1}{c}{114.4}\\
P-O$^{({\rm B^{\prime \prime}})}$& \multicolumn{1}{c|}{1.588} & O$^{({\rm A^{\prime \prime}})}$-P-O$^{({\rm C^{\prime \prime}})}$ & \multicolumn{1}{c}{115.0} \\
P-O$^{({\rm C^{\prime \prime}})}$& \multicolumn{1}{c|}{1.594} & O$^{({\rm A^{\prime \prime}})}$-P-O$^{({\rm D^{\prime \prime}})}$ & \multicolumn{1}{c}{116.5} \\
P-O$^{({\rm D^{\prime \prime}})}$& \multicolumn{1}{c|}{1.599} &  \\
\hline \hline
\end{tabularx}
\label{table:bonds-angles}
\end{table}

\begin{figure}[htb]
\centering
    \includegraphics*[trim=120pt 250pt 120pt 250pt, width=8.6cm]{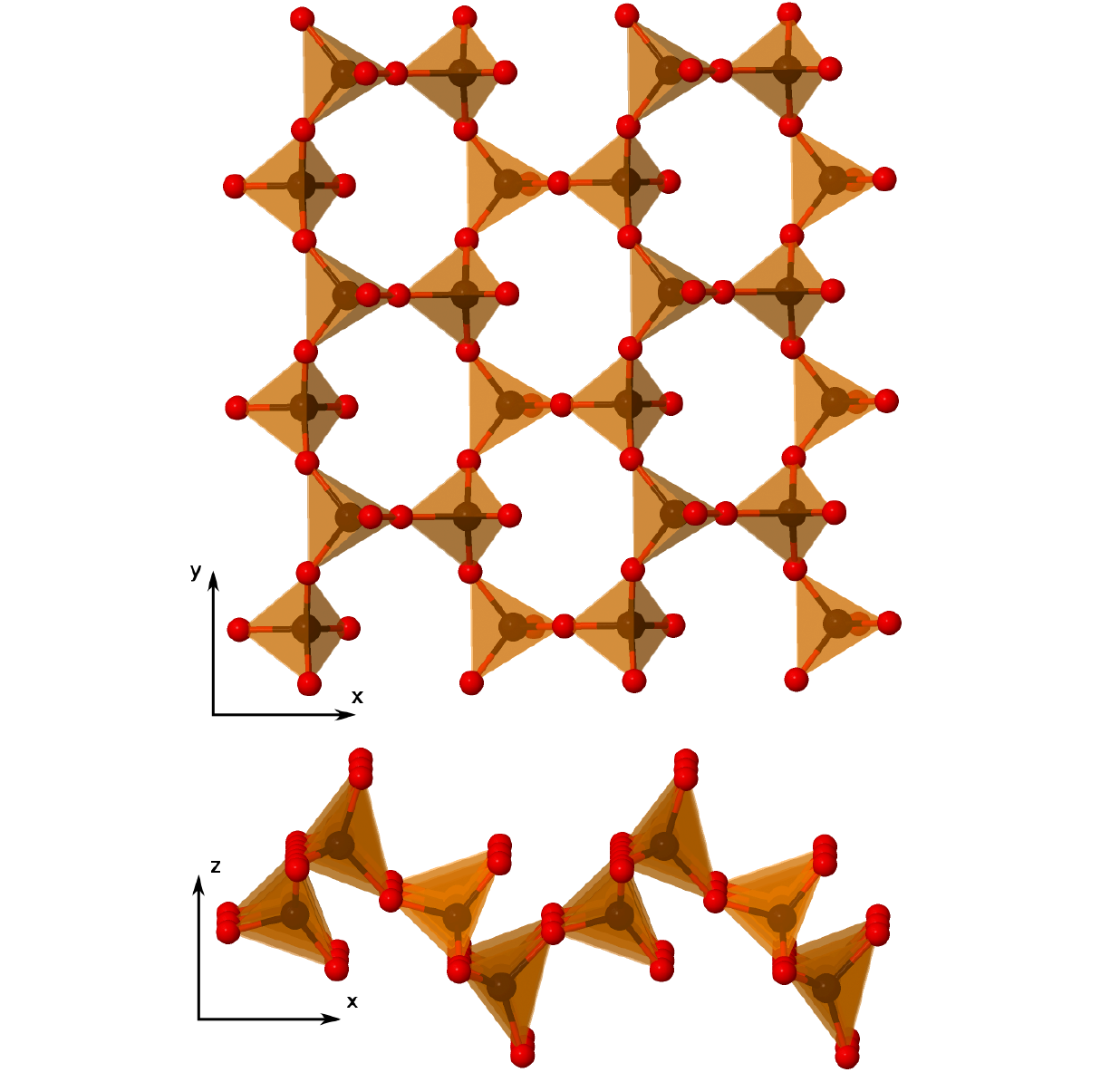}
 \caption{\small Structure of single layer o$^\prime$-P$_2$O$_{5}$. }
\label{fig:Op-p4o10}
\end{figure}

The planar $p$-P$_4$O$_{10}$ structure is entirely different from the layered o$^\prime$-P$_2$O$_5$ form 
previously reported by several experimental studies,\cite{stachel1995} shown in Fig.\ref{fig:Op-p4o10}.
While both consist of rings of six PO$_4$ tetrahedra, in o$^\prime$-P$_2$O$_5$ the tetrahedra are organised in alternating rows along the [100].
Further, in profile o$^\prime$-P$_2$O$_5$ resembles a sawtooth, whereas $p$-P$_4$O$_{10}$ has the same waved appearence as phosphorene.

Nearly degenerate in energy with the planar $p$-P$_4$O$_{10}$ form (Fig.\ref{fig:eb}), we find the tubular $t$-P$_4$O$_{10}$ structure, shown in Fig.\ref{fig:t-p4o10}. The $t$-P$_4$O$_{10}$ structure is a tubular (1D) form consisting of eight-membered rings, connected to each other at an angle of approximately 100$^{\circ}$ (see for example the angle between the D$^{\prime}$ and B$^{\prime}$ bonds or D$^{\prime \prime}$ and C$^{\prime \prime}$ bonds in Fig.\ref{fig:t-p4o10}), these rings consisting of alternated phosphorus and bridging oxygen atoms. Only inversion symmetry is present, hence the space group is P$^{-1}$ (C$_i^1$).
There are two inequivalent (although very similar) PO$_4$ units, depicted in Fig.\ref{fig:t-p4o10} (left side). 
The bond lengths and angles are listed in Table \ref{table:bonds-angles}.

Similar to $p$-P$_4$O$_{10}$, $t$-P$_4$O$_{10}$ can be seen as a corner-linked network of PO$_4$ tetrahedra.
However, different from polyphosphate chains,\cite{greenwood1997} in these tubular forms each tetrahedra shares three vertices,
forming a tubular structure rather than a simple chain structure.

For intermediate concentrations, we find two metastable phosphorene oxides in which oxygen is present only on the phosphorene surface, in the form of dangling oxygen atoms. We refer to these as surface forms, and in particular as $s$-P$_4$O$_2$ for the single-surface oxidized, and $s$-P$_4$O$_4$ for the double-surface oxidized; the two structures are shown in Fig.\ref{fig:structures} panel (c).
As a result of oxygen chemisorption, the planes containing the top and the bottom zigzag ridges are not parallel to each other anymore (as in pristine phosphorene). The two P atoms in the same zigzag ridge are shifted from each other by about 0.5~\AA\ in the direction perpendicular to the surface.
In contrast with the other planar forms, the lattice deformation in this case is minimal.

\subsection{Stability}

\begin{figure}[htb]
\centering
    \includegraphics*[trim=10pt 190pt 15pt 100pt, width=8.6cm]{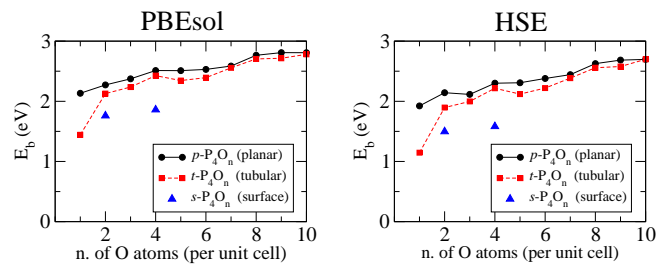}
 \caption{\small Average binding energy E$_b$ per oxygen atom as a function of oxygen concentration calculated with the PBEsol and HSE functionals. Similar results are found for the PBE functional (not shown). Black circles indicate planar PO forms ($p$-P$_4$O$_{n}$), red squares tubular forms ($t$-P$_4$O$_{n}$) and blue triangles surface forms ($s$-P$_4$O$_{n}$). The lines are a guide for the eye. }
\label{fig:eb}
\end{figure}

The stability of the POs can be quantified using the binding energy per oxygen atom, defined as
\begin{align}
E_b = -\dfrac{1}{N_{\rm O}} \left[ E_{ox} - \left(E_p+\dfrac{1}{2}N_{\rm O} E_{\rm O_2}\right)\right]
\label{eq:e-bind}
\end{align}
where $N_{\rm O}$ is the number of O atoms in the unit cell, $E_{ox}$, $E_p$ and $E_{{\rm O}_2}$ are the total energies of the PO, the pristine phosphorene, and the O$_2$ (triplet) molecule, respectively. 
From the definition above, a value of $E_b>0$ indicates that the oxide formation is energetically favored in the presence of O$_2$.
The average binding energies per O atom E$_b$ for the most stable tubular, planar and surface POs calculated with the PBEsol and HSE functionals are shown in Fig.\ref{fig:eb}. 
The gain in energy due to oxygen chemisorption in phosphorene is very large across the whole concentration range (from $\sim$1.8 to 2.9~eV per O atom).

Regardless of the oxygen concentration, we invariably find that the lowest energy structure is a planar ($p$-P$_4$O$_{n}$) phosphorene oxide,
with the tubular forms lying slightly higher in energy.   
In addition, for low and medium oxygen concentrations we find a manifold of states corresponding to different oxygen arrangements
that have nearly the same energy.
Obviously, for P$_4$O$_{10}$ the phosphorene lattice is saturated with oxygen, so only two new structures were found ($p$-P$_4$O$_{10}$ and $t$-P$_4$O$_{10}$).

For the lowest concentration considered, $n$=1, the chemisorption of dangling oxygen is favored relative to bridging oxygens; in fact, $p$-P$_4$O$_{1}$ has a chemisorbed dangling oxygen (E$_b$=2.1~eV), while on $t$-P$_4$O$_{1}$ we have a bridging oxygen (E$_b$=1.4~eV). 
However, for all O concentrations higher than $n$=1, no clear difference emerges in the energetics of dangling or bridging oxygens. The amount of energy gained after O chemisorption is dictated by the interplay between Coulomb repulsion of localized charges on the phosphorene surface and strain interactions, the former due to dangling oxygens and the latter mainly governed by bridging oxygens. Since bridging oxygens increase the P-P distance by about 0.7~\AA\ (Fig.\ref{fig:motifs}b), the insertion of oxygen bridges between neighboring dangling oxygens can substantially reduce their Coulomb repulsion, greatly stabilizing the PO.
This explains the counter-intuitive increase in binding energy per oxygen atom with increasing oxygen concentration.

\begin{figure}[htb]
\centering
    \includegraphics*[trim=0pt 285pt 0pt 290pt, width=8.6cm]{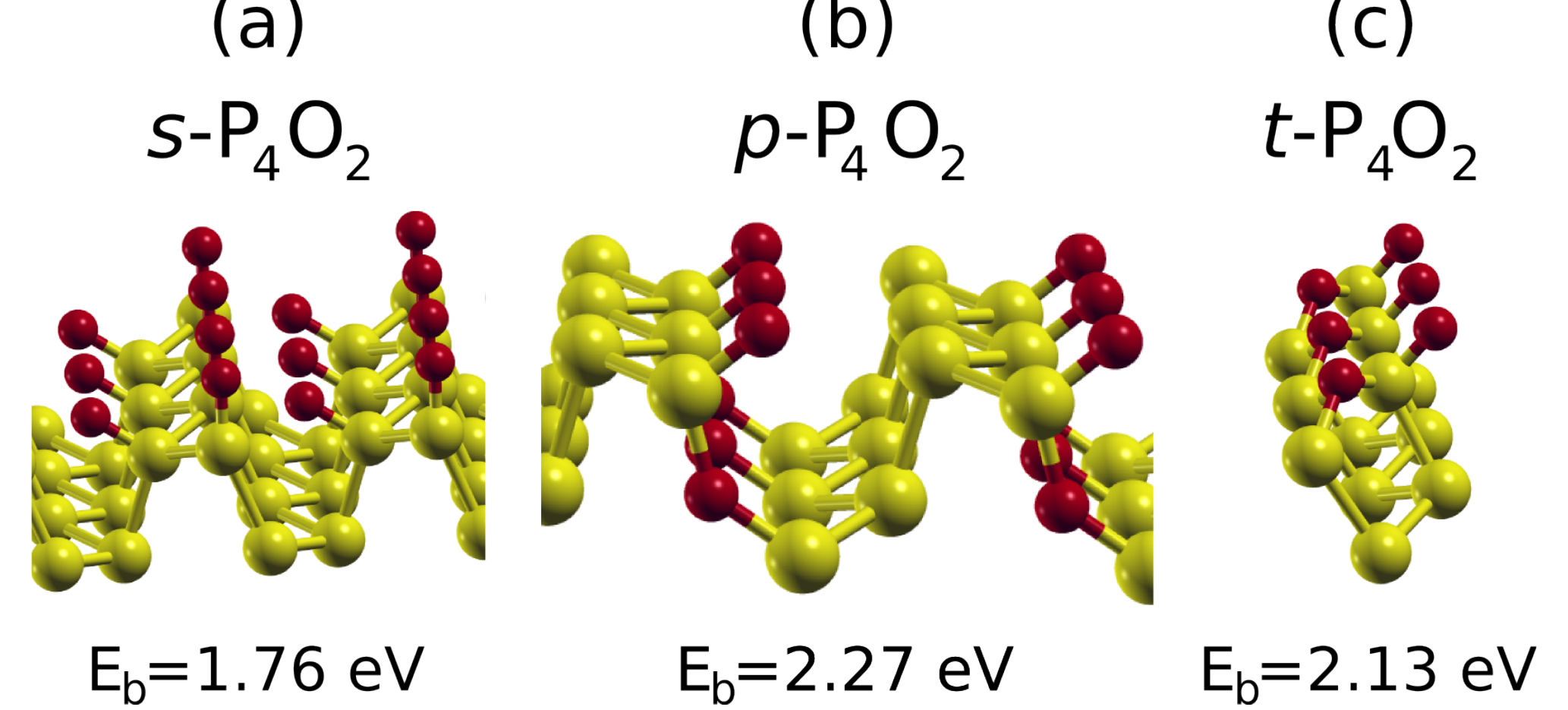}
 \caption{\small Different forms of phosphorene oxides and their average binding energy per oxygen atom E$_b$ ($n$=2). }
\label{fig:eb-n=2}
\end{figure}

The effect of the reduction of Coulomb repulsion due to the insertion of oxygen bridges is particularly evident for the case $n$=2, shown in Fig.\ref{fig:eb-n=2}. Even though for $n$=1 dangling oxygen chemisorption is favored over bridging oxygen chemisorption, when two dangling oxygens are chemisorbed on the same side of the phosphorene surface their strong Coulomb repulsion raises the energy of the compound, resulting in E$_b$=1.76~eV for the surface form $s$-P$_4$O$_{2}$ (blue triangle at $n$=2 in Fig.\ref{fig:eb}). This energy can be reduced by forming oxygen bridges instead of dangling oxygen, 
and therefore there are numerous other $p$-P$_4$O$_{2}$ and $t$-P$_4$O$_{2}$ forms containing bridging oxygen that are more stable
(up to 1.02~eV and 0.74~eV, respectively) than $s$-P$_4$O$_{2}$.
Similar considerations apply to the double-side surface oxidized form $s$-P$_4$O$_{4}$, which is $\sim$1.8~eV and $\sim$1.4~eV higher in energy than the most stable planar and tubular forms, respectively, with the same concentration. From these simple considerations, we can argue that in near-equilibrium conditions the surface forms $s$-P$_4$O$_{2}$ and $s$-P$_4$O$_{4}$ will not be the primary outcome of phosphorene oxidation, contrary to what has been recently proposed\cite{wang}.
Nevertheless,  these surface oxidized forms can possibly be favored by specific kinetic factors. Still, since the activation energies for dangling oxygen formation and for oxygen insertion (bridges formation) are very similar (0.54 and 0.69 eV, respectively),\cite{ziletti2014} it is unlikely that these surface oxide phases will form even at low temperatures.

We also found that the other surface forms proposed in the literature\cite{daia2014} are not stable at both PBE and PBEsol level, and after lattice relaxation and geometry optimization (with a tight convergence threshold for forces of 10$^{-3}$~eV/\AA) the two forms presented here, $s$-P$_4$O$_{2}$ and $s$-P$_4$O$_{4}$, are found.

For the maximum oxygen concentration ($n$=10), the $p$-P$_4$O$_{10}$ and $t$-P$_4$O$_{10}$ forms have nearly degenerate energies, with E$_b$=2.81~eV (2.70~eV) and E$_b$=2.78~eV (2.70~eV) at the PBEsol (HSE) level, respectively. 
They are nearly degenerate in energy with the o$^\prime$ form, which has a binding energy of 2.81~eV with PBEsol (2.68~eV with HSE) in the monolayer form,
or 2.88~eV (at the PBEsol level) in the bulk form.
They also are the most stable POs among all O concentrations considered in this work.
Their structures, which are essentially a homogenous network of three-corner linked PO$_4$ tetrahedra (Fig.\ref{fig:p-p4o10} and Fig.\ref{fig:t-p4o10}), minimize both strain interactions due to oxygen bridges and Coulomb repulsion from dangling oxygens.  

\subsection{Electronic properties}

\begin{figure}[htb]
\centering
    \includegraphics*[trim=8pt 60pt 15pt 182pt, width=8.6cm]{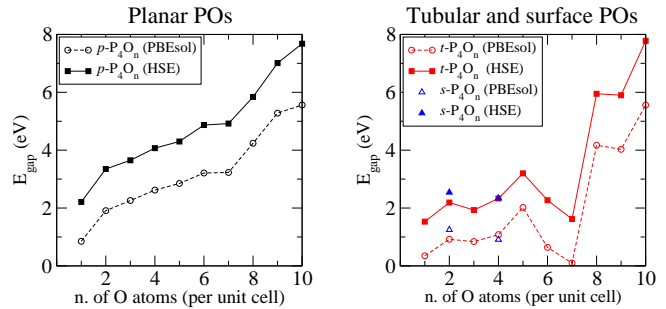}
 \caption{\small Bandgap energy E$_{{\rm gap}}$ calculated with the PBEsol and HSE functionals for planar (left panel), tubular and surface forms (right panel) as a function of oxygen concentration. The lines are a guide for the eye.}
\label{fig:egap}
\end{figure}

Chemisorption of oxygen atoms and the formation of POs drastically changes the electronic properties of phosphorene. The bandgap E$_{{\rm gap}}$ of both planar and tubular oxides is plotted as a function of oxygen concentration in Fig.\ref{fig:egap}. Both planar and tubular phosphorene oxides are found to be semiconducting or insulating, depending on the O concentration. 

The bandgap of planar POs (black squares) increases monotonically with oxygen concentration, from  2.21~eV ($n=1$) to 7.69~eV ($n=10$) at the HSE level.
Note that the bandgaps calculated at the PBEsol level follow almost precisely the same trends, but are simply shifted to smaller gaps compared to HSE by roughly 1.5~eV across the entire O concentration range (see Fig.\ref{fig:egap}).

\begin{figure}[htb]
\centering
    \includegraphics*[trim=10pt 150pt 10pt 180pt, width=8.1cm]{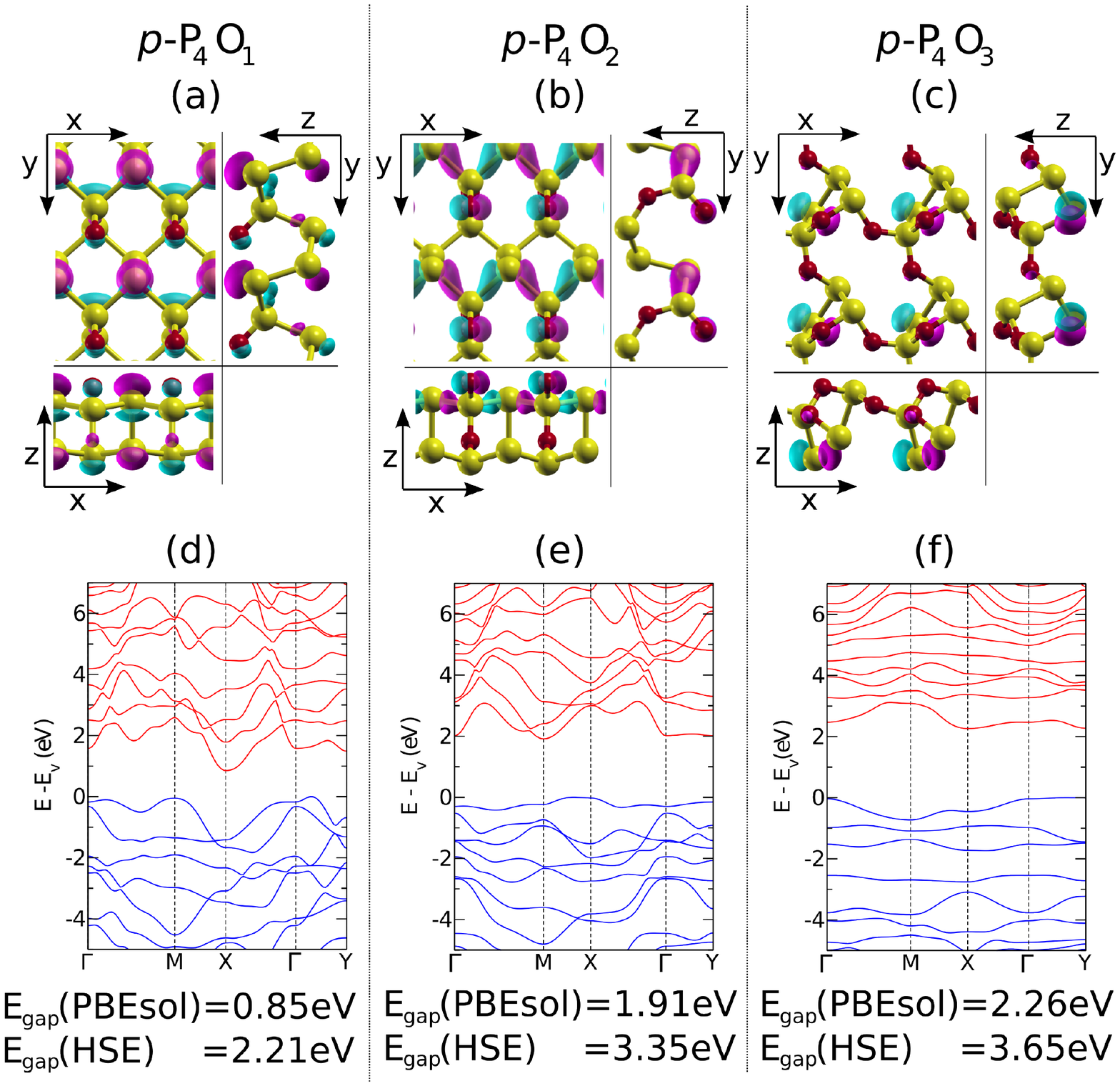}
 \caption{\small (a)(b)(c) Atomic structure projections and valence band isosurface for low concentration planar POs: $p$-P$_4$O$_{1}$ (a), $p$-P$_4$O$_{2}$ (b), $p$-P$_4$O$_{3}$ (c). The PBEsol electronic band structures are shown in (d) (e) (f), respectively. The top of the valence band is set to zero. The energy bandgaps at the PBEsol and HSE level are also given.}
\label{fig:wf-conf}
\end{figure}

The increase in bandgap with O concentration in planar POs is due to the increasingly ionic character of the bonds,
and the resulting wavefunction localization. The series $p$-P$_4$O$_{n}$ ($n$=1,2,3) is a particularly explanatory example of this phenomenon, and it is shown in Fig.\ref{fig:wf-conf}. In $p$-P$_4$O$_{1}$ one dangling oxygen is chemisorbed, corresponding to a 25\% concentration. The band structure is strongly modified compared to pristine phosphorene (shown in \emph{Supplemental Material}), but the valence band is still delocalized over all the crystal, as shown in Fig.\ref{fig:wf-conf}a, and still exhibits moderate dispersion. The bandgap is now indirect, but its value is still close to pristine phosphorene (2.21~eV \emph{vs.} 1.77~eV of pristine phosphorene at the HSE level in our calculation).
If more oxygen atoms are added, the band structure drastically changes (Fig.\ref{fig:wf-conf}e-\ref{fig:wf-conf}f). 
In $p$-P$_4$O$_{2}$, the valence band becomes localized on the top zigzag ridge (Fig.\ref{fig:wf-conf}b).
The valence band is now nearly completely flat, and the conduction band generally has less dispersion than that of $p$-P$_4$O$_{1}$. The bridging oxygen thus effectively creates nanoribbons, and strongly increases the bandgap (3.35~eV with HSE). 
In $p$-P$_4$O$_{3}$ (Fig.\ref{fig:wf-conf}c-\ref{fig:wf-conf}f) additional oxygen bridges, intra-ridge and inter-ridge, are formed. The valence band is now localized on P, and both valence and conduction bands have very low dispersion due to their localization. The bandgap is 3.65~eV at the HSE level. 
For $n$$>$3, all $p$-PO band structures preserve the low dispersion present in $p$-P$_4$O$_{3}$, together with an increasing bandgap. 
Their atomic and electronic band structures are shown in \emph{Supplemental Material}.

Tubular phosphorene oxides are also insulating, with bandgaps covering a large range of energies from 1.62~eV to 7.78~eV (at HSE level). 
In contrast to what we have described for planar POs, we did not find that tubular POs exhibit a monotonic increase in bandgap with oxygen concentration (see Fig.\ref{fig:egap}). We can separately consider two regimes, $n\leq7$ and $n>7$.

\begin{figure}[htb]
\centering
    \includegraphics*[trim=5pt 180pt 10pt 210pt, width=8.6cm]{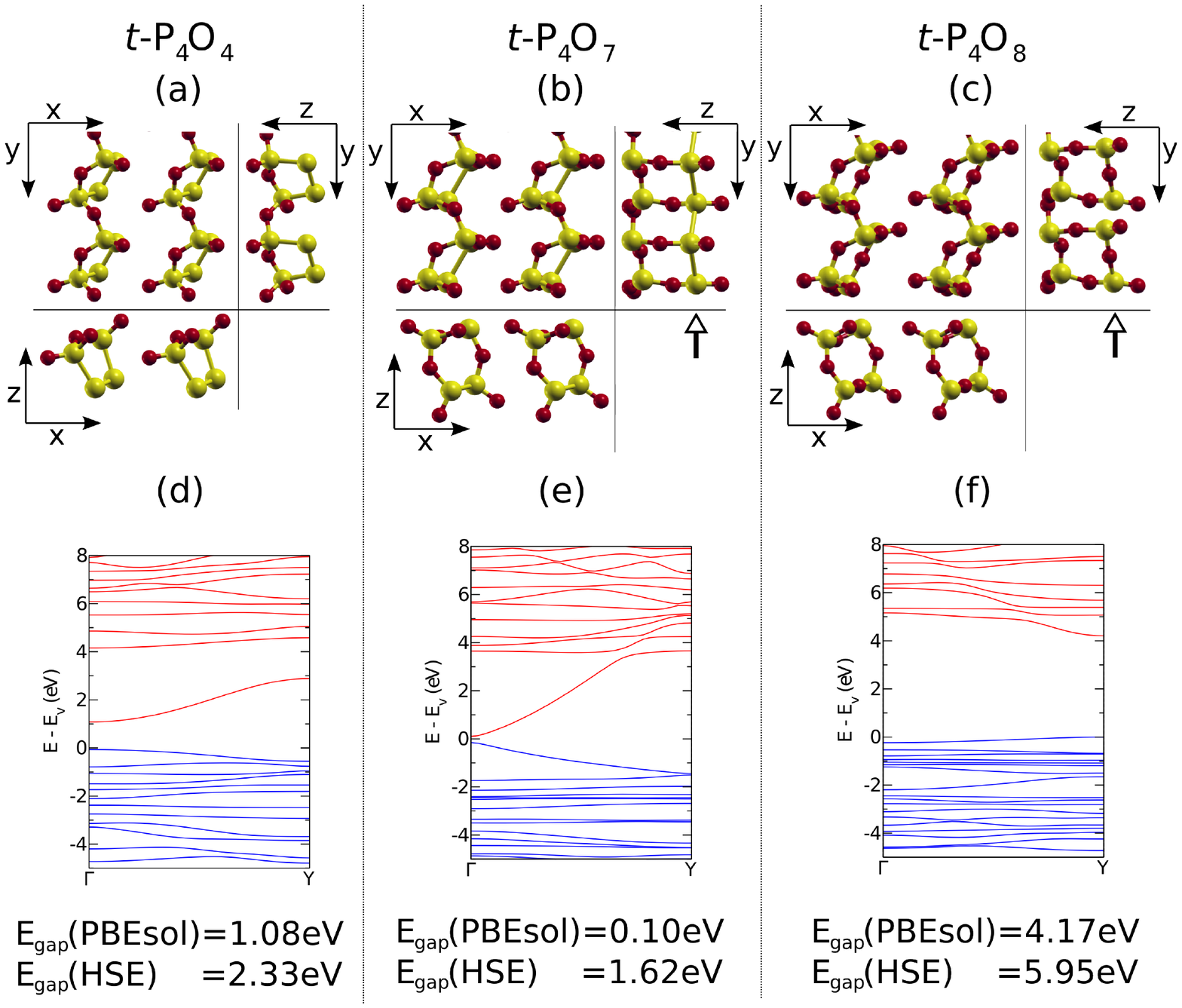}
 \caption{\small (a)(b)(c) Atomic structure projections and conduction band isosurface for representative tubular POs: $t$-P$_4$O$_{4}$ (a), $t$-P$_4$O$_{7}$ (b), $t$-P$_4$O$_{8}$ (c). The PBEsol electronic band structures are shown in (d) (e) (f), respectively. Only the dispersion along the 1D chains is shown. The top of the valence band is set to zero. The energy bandgaps at the PBEsol and HSE level are also given.}
\label{fig:tub-bands}
\end{figure}

The atomic structure and electronic band structure of some representative tubular POs are shown in Fig.\ref{fig:tub-bands}.
The compound $t$-P$_4$O$_{4}$ (Fig.\ref{fig:tub-bands}a) presents a feature common to all low and medium concentration tubular POs ($n\leq7$): both occupied and unoccupied states around the Fermi energy are nearly flat, while the conduction band is dispersive and well-separated from the manifold of occupied (below) and empty states (above). The conduction band in this particular case has $\sim$2~eV bandwidth in all directions ($\Gamma$-M, M-X and $\Gamma$-Y), but it is flat on the X-$\Gamma$ direction, which is the direction perpendicular to the chain composing this tubular PO; this is the signature of the 1D nature of tubular POs. The strong Coulomb repulsion between dangling oxygens separates the chains, which interact only through (weak) van der Waals forces.  

The compound $t$-P$_4$O$_{7}$ has a similar band structure, but shows increased dispersion of the conduction band (Fig.\ref{fig:tub-bands}b-\ref{fig:tub-bands}e).
 The conduction band arises from delocalized phosphorus $p$ orbitals at the bottom of the PO layer, where no oxygen bridges are present (see right side of Fig.\ref{fig:tub-bands}b). The bandgap is only 1.62~eV (HSE level).

In the high concentration regime ($n\geq8$), the scenario is quite different.
Chemisorption of a bridging oxygen atom in the phosphorus chain (marked by an arrow in Fig.\ref{fig:tub-bands}b,c) breaks the conjugation, localizing both conduction and valence bands, and greatly increasing E$_{gap}$. This is particularly evident in $t$-P$_4$O$_{8}$ (Fig.\ref{fig:tub-bands}c-\ref{fig:tub-bands}f): the conduction band, in contrast to the situation discussed above, is now very close to the manifold of unoccupied states, and presents no dispersion. The bandgap increases by roughly $\sim$4~eV at both PBEsol and HSE levels, making the material an insulator. 
The surface forms $s$-P$_4$O$_{2}$ and $s$-P$_4$O$_{4}$ are instead both semiconducting with HSE bandgaps of 2.54~eV and 2.34~eV, respectively. 

\begin{figure}[htb]
\centering
    \includegraphics*[trim=10pt 120pt 10pt 125pt, width=8.6cm]{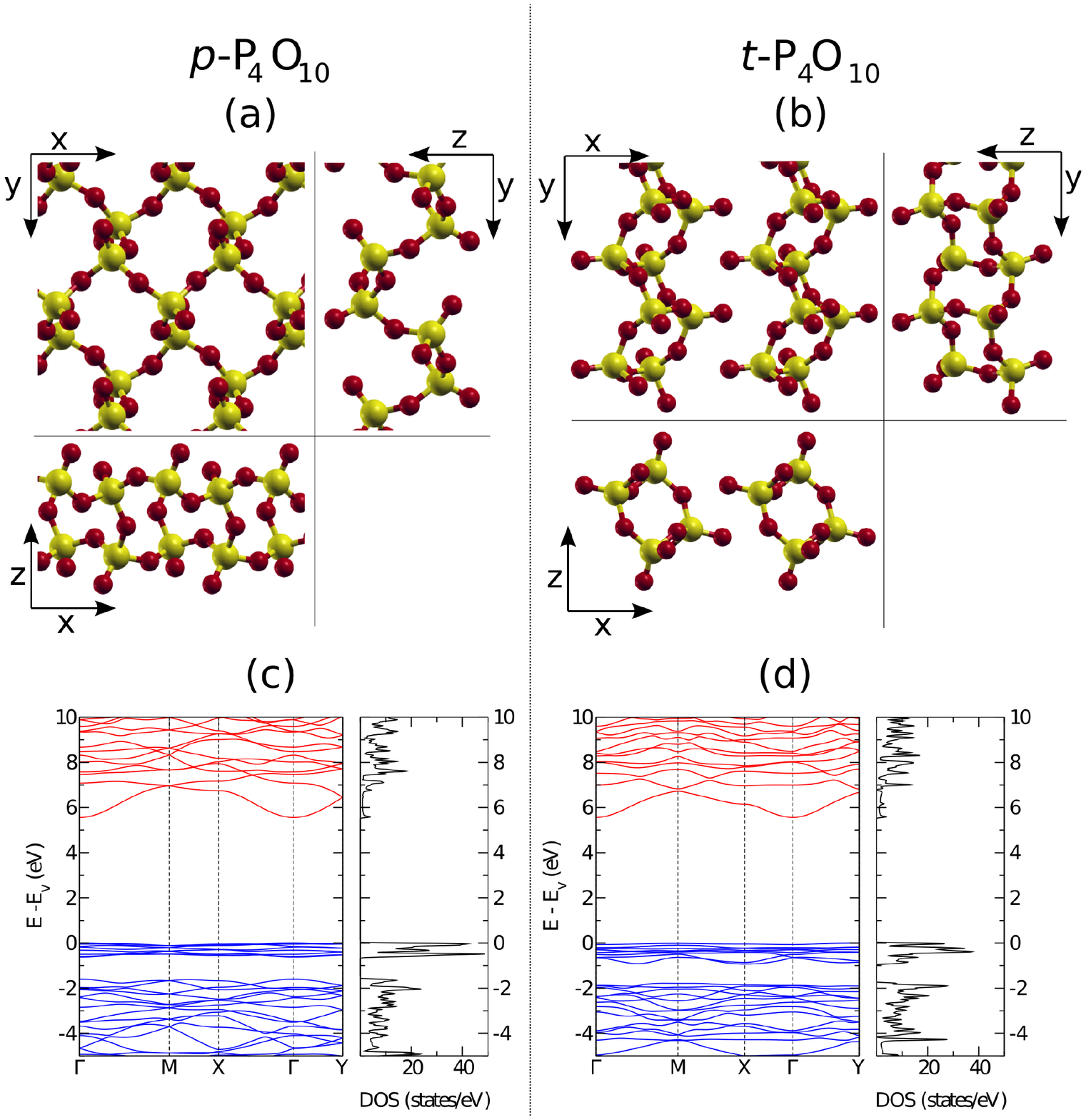}
 \caption{\small (a)(b)Atomic structure projections for maximum oxidation POs: $p$-P$_4$O$_{10}$ (a) and $t$-P$_4$O$_{10}$ (b). The PBEsol electronic band structures are shown in (c) (d), respectively. The top of the valence band is set to zero.}
\label{fig:max-oxid}
\end{figure}
Electronic band structure and density of states (DOS) of the maximally oxidized phosphorene oxides $p$-P$_4$O$_{10}$ and $t$-P$_4$O$_{10}$ are shown in Fig.\ref{fig:max-oxid}. 
Both POs are insulators, and they have remarkably similar band structures. There is a manifold of 8 nearly degenerate dispersion-less occupied states between -1~eV and 0~eV for both species; these states are $p$ orbitals localized on dangling oxygens, bridging oxygens or both, and they do not have any appreciable electron density on the P atoms. The conduction band consists instead of $p$ orbitals on dangling oxygens and on P-O bridges, and it is slightly dispersive. The bandgap is 5.56~eV for both species at the PBEsol level, enlarged at 7.68~eV for $p$-P$_4$O$_{10}$ and at 7.78~eV for $t$-P$_4$O$_{10}$ with the HSE functional. Interestingly, we find that monolayer o$^\prime$-P$_2$O$_5$ has a similar HSE bandgap (7.45~eV).

For pristine phosphorene and the maximally oxidized phosphorene oxides we calculated the bandgaps also within the GW approximation.\cite{hedin1965,hybertsen1986} For the pristine phosphorene we obtain 1.70~eV, close to the 1.60~eV of Ref.\cite{rudenko2014}, but different by 0.3~eV from Tran {\em et al.}\cite{tran2014} We note that all the convergence parameters used in this work are higher than those used in these previous studies (see \emph{Supplemental Material} for a complete comparison).  

For $p$-P$_4$O$_{10}$ and $t$-P$_4$O$_{10}$ we obtain GW bandgaps of 8.50~eV and 8.69~eV, respectively, close to the HSE results. 
The bandgaps for pristine phosphorene and maximal oxidation POs calculated with different levels of theory are summarised in Table \ref{table:egap}.

\begin{table}[htb]
\caption{Energy bandgaps for pristine phosphorene and maximal oxidation POs.} 
\centering 
\begin{tabular}{l c  c  c } 
\hline\hline 

\multirow{2}{*}{System} & & E$_{\rm gap}$(eV) & \\
 & PBEsol & HSE & GW\\
\hline

Phosphorene \hspace{1em} &  0.72   & 1.77 & 1.70    \\  

$p$-P$_4$O$_{10}$ & 5.56 & 7.68 & 8.50   \\  
				
$t$-P$_4$O$_{10}$ & 5.56 & 7.78 &  8.69    \\  
			       
\hline\hline 
\end{tabular}
\label{table:egap}
\end{table}

\subsection{Vibrational properties}
\label{subsec:vibrational-properties}

\begin{figure}[htb]
\centering
    \includegraphics*[trim=65pt 0pt 70pt 0pt, width=8.6cm]{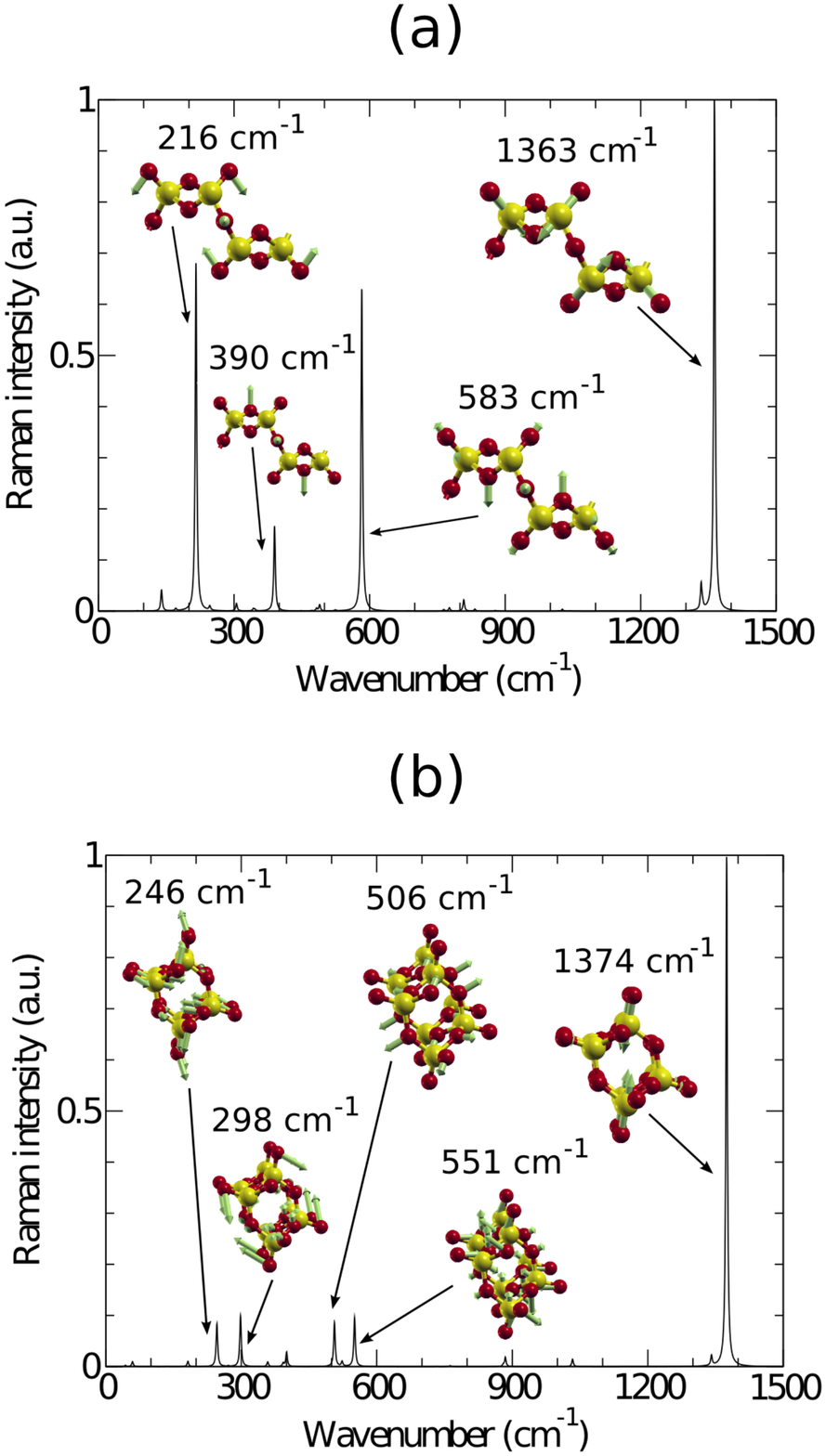}
 \caption{\small Raman spectra and corresponding Raman active modes for $p$-P$_4$O$_{10}$ (a) and $t$-P$_4$O$_{10}$ (b). The length of the arrows is proportional to the amplitude of motion. }
\label{fig:raman}
\end{figure}

Even though the Raman spectrum of POs changes quite significantly depending on the oxygen concentration, it is possible to identify three distinctive vibrational regions common throughout the whole P$_4$O$_n$ series. Some of the typical vibrational modes relevant in these different regions are presented in Fig.~\ref{fig:raman}. 
The low frequency region of the vibrational spectra ($<$300~cm$^{-1}$) is dominated by bending modes of the P=O groups relative to the rest of the structure, but also includes a small component of P-O-P bending. The presence of P-O-P bending modes increases with frequency, and becomes strongly predominant in the region between 500 and 1000~cm$^{-1}$. For the surface forms ($s$-P$_4$O$_{2}$ and $s$-P$_4$O$_{4}$) in which only dangling oxygens are present, there are indeed no vibrational modes between 500~cm$^{-1}$ and 1000~cm$^{-1}$. 
In the high frequency region, the vibrational spectrum instead only involves P=O stretching modes; these modes are decoupled from the other vibrations because the P=O bonds are much stronger than P-O bridges, as in the case of the molecular (bulk) phosphorus oxides\cite{valentim1997,valentim1998}. The P=O stretching modes are separated from the other modes by at least 200~cm$^{-1}$ for all oxygen concentrations, and there are as many as the number of dangling oxygen in the PO. This observation opens the possibility of direct experimental detection of the number of dangling oxygens through Raman spectroscopy. 
P=O stretching modes start at 1063~cm$^{-1}$ for $n$=1 ($p$-P$_4$O$_{1}$) and monotonically blueshift with increasing oxygen concentration till about $\sim$1370~cm$^{-1}$ for maximum oxidation. This is consistent with the shortening of P=O bond found with increasing oxygen linkage, as shown in Table \ref{table:blength} and analogous to the observed blueshift of P=O vibrational frequencies in the series P$_4$O$_{m}$ ($m$=6$-$10) measured in solid argon\cite{mielke1989}. 

Raman spectra and Raman active modes of $p$-P$_4$O$_{10}$ and $t$-P$_4$O$_{10}$ are shown in Fig.\ref{fig:raman}a and \ref{fig:raman}b, respectively.  
In $p$-P$_4$O$_{10}$, there are three high intensity Raman peaks at 216~cm$^{-1}$, 583~cm$^{-1}$ and 1363~cm$^{-1}$ and a low intensity peak at 390~cm$^{-1}$.
The first peak (216~cm$^{-1}$) consists of the concerted wag mode of the four P=O dangling oxygens (scissoring), together with an symmetric P-O-P stretch of the two inter-ridge bridges (purple in Fig.\ref{fig:p-p4o10}b). The second strong peak at 583~cm$^{-1}$ is a combination of the P-O-P symmetric stretch of the middle intra-ridge bridges (brown in Fig.\ref{fig:p-p4o10}b) and the P=O stretch of all dangling oxygens. The last strong peak (1363~cm$^{-1}$) involves the stretching of the P=O bonds.
The low intensity peak at 390 cm$^{-1}$ is instead the P-O-P symmetric stretching of the lower and higher intra-ridge bridges (grey in Fig.\ref{fig:p-p4o10}b).

In contrast, only one strong peak at 1374 cm$^{-1}$, corresponding to the P=O stretching of two dangling oxygens, is present in the Raman spectrum of $t$-P$_4$O$_{10}$, as shown in Fig.\ref{fig:raman}b.  Four low intensity peaks, however, appear in the range 250-550~cm$^{-1}$. 

The mode at 246 cm$^{-1}$ involves P=O bond scissoring with a contribution from an asymmetrical bending of the P-O-P bridges, while the one at 298 cm$^{-1}$ involves P=O scissoring such that the top and bottom layer dangling oxygen moves towards the tubular oxide center. Finally, at 506 cm$^{-1}$ we find a ring-breathing mode (P-O-P symmetrical bending of inter-ridge bridges) and at 551 cm$^{-1}$ another ring breathing mode plus P=O stretching. 

Raman and IR spectra for planar and tubular forms for all oxygen concentrations are reported in \emph{Supplemental Material}. 

\section{Conclusions}
We have described a family of two dimensional phosphorus oxides obtained by oxidation of phosphorene, with oxygen content up to 10 oxygen per phosphorene unit cell
(P$_2$O$_5$). 

The gain in binding energy due to oxygen chemisorption in phosphorene is very large (from $\sim$1.8 to 2.9~eV per O atom),
and increases with the oxygen concentration. 
In parallel with the two-dimensional, planar forms, there are tubular (polymeric) forms with similar formation enthalpy.
The planar forms are lower in energy than the tubular forms for all O concentrations; the energy difference is however quite small (typically$<$0.1~eV per O atom), hence both forms are likely to coexist under normal experimental conditions, forming ordered and disordered (amorphous) domains depending on the oxidation process, local impurities and defects. 
It is also feasible that different forms of POs can be experimentally created by a suitable choice of growth conditions, in much the same way as the (bulk) series P$_4$O$_{6+m}$ (0$\leq$$m$$\leq$4) can be obtained by varying the atmosphere, temperature and pressure in which P$_4$ is treated\cite{jansen1981,greenwood1997}. 
Moreover, both planar and tubular forms have very similar formation enthalpy of the thermodynamically most stable known phosphorus pentoxide (o$^\prime$-P$_2$O$_5$).

The phosphorene oxides can be used as a transparent tunneling materials, namely on black-phosphorus or phosphorene-based devices.
The bandgap energy of planar P$_4$O$_n$ increases with $n$, and is 8.5~eV for $n=10$.
The tubular forms are also insulating, and close to the oxygen saturation limit (P$_2$O$_5$),
have nearly the same gap energy, and therefore their coexistence with the planar form does not jeopardize its transparency or electrical insulation properties.

Though the planar and the tubular form of P$_2$O$_5$ give rise to high frequency Raman scattering peaks 
with very close energies (at 1363 and 1374 cm$^{-1}$), the planar form can be identified by an intense peak at 583 cm$^{-1}$,
just above the Raman edge of black phosphorus (470 cm$^{-1}$ at room temperature\cite{sugai1985}).

\section*{Acknowledgments}
A.Z. and D.F.C. acknowledge NSF grant CHE-1301157 and also an allocation of computational resources from Boston University's Office of Information Technology and Scientific Computing and Visualization. A.Z. also acknowledges the support from NSF grant CMMI-1036460, Banco Santander. A.H.C.N., A.C. and P.E.T.  acknowledge the National Research Foundation, Prime Minister's Office, Singapore, under its Medium Sized Centre Programme and CRP award "Novel 2D materials with tailored properties: beyond graphene" (R-144-000-295-281).

\bibliography{citations-long}{}

\begin{thebibliography}{57}
\providecommand{\natexlab}[1]{#1}
\providecommand{\url}[1]{\texttt{#1}}
\expandafter\ifx\csname urlstyle\endcsname\relax
  \providecommand{\doi}[1]{doi: #1}\else
  \providecommand{\doi}{doi: \begingroup \urlstyle{rm}\Url}\fi

\bibitem[Rodin et~al.(2014)Rodin, Carvalho, and Castro~Neto]{rodin2014}
S.~Rodin, A.\, A.~Carvalho, and H.~Castro~Neto, A.\.
\newblock Strain-induced gap modification in black phosphorus.
\newblock \emph{Phys. Rev. Lett.}, 112:\penalty0 176801, May 2014.
\newblock \doi{10.1103/PhysRevLett.112.176801}.
\newblock URL \url{http://link.aps.org/doi/10.1103/PhysRevLett.112.176801}.

\bibitem[Tran et~al.(2014)Tran, Soklaski, Liang, and Yang]{tran2014}
Vy~Tran, Ryan Soklaski, Yufeng Liang, and Li~Yang.
\newblock Layer-controlled band gap and anisotropic excitons in few-layer black
  phosphorus.
\newblock \emph{Phys. Rev. B}, 89:\penalty0 235319, Jun 2014.
\newblock \doi{10.1103/PhysRevB.89.235319}.
\newblock URL \url{http://link.aps.org/doi/10.1103/PhysRevB.89.235319}.

\bibitem[Wei and Peng(2014)]{wei2014}
Qun Wei and Xihong Peng.
\newblock Superior mechanical flexibility of phosphorene and few-layer black
  phosphorus.
\newblock \emph{Applied Physics Letters}, 104\penalty0 (25):\penalty0 251915,
  2014.
\newblock \doi{http://dx.doi.org/10.1063/1.4885215}.
\newblock URL
  \url{http://scitation.aip.org/content/aip/journal/apl/104/25/10.1063/1.4885215}.

\bibitem[Koenig et~al.(2014)Koenig, Doganov, Schmidt, Castro~Neto, and
  Ozyilmaz]{koenig2014}
Steven~P. Koenig, Rostislav~A. Doganov, Hennrik Schmidt, A.~H. Castro~Neto, and
  Barbaros Ozyilmaz.
\newblock Electric field effect in ultrathin black phosphorus.
\newblock \emph{Applied Physics Letters}, 104\penalty0 (10):\penalty0 103106,
  2014.
\newblock \doi{http://dx.doi.org/10.1063/1.4868132}.
\newblock URL
  \url{http://scitation.aip.org/content/aip/journal/apl/104/10/10.1063/1.4868132}.

\bibitem[Xia et~al.()Xia, Wang, and Jia]{xia2014}
Fengnian Xia, Han Wang, and Yichen Jia.
\newblock Rediscovering black phosphorus: A unique anisotropic 2d material for
  optoelectronics and electronics.
\newblock \emph{arXiv:1402.0270}.

\bibitem[Buscema et~al.(2014)Buscema, Groenendijk, Blanter, Steele, van~der
  Zant, and Castellanos-Gomez]{buscema2014}
Michele Buscema, Dirk~J. Groenendijk, Sofya~I. Blanter, Gary~A. Steele, Herre
  S.~J. van~der Zant, and Andres Castellanos-Gomez.
\newblock Fast and broadband photoresponse of few-layer black phosphorus
  field-effect transistors.
\newblock \emph{Nano Letters}, DOI: 10.1021/nl5008085, 2014.
\newblock \doi{10.1021/nl5008085}.
\newblock URL \url{http://pubs.acs.org/doi/abs/10.1021/nl5008085}.

\bibitem[A.~Ziletti and Neto()]{ziletti2014}
D.~K. Campbell D. F.~Coker A.~Ziletti, A.~Carvalho and A.~H.~Castro Neto.
\newblock Oxygen defects in phosphorene.
\newblock \emph{arXiv:1407.5880}.

\bibitem[Doganov et~al.()Doganov, OÕFarrell, Koenig, Yeo, Watanabe, Taniguchi,
  Ziletti, Carvalho, Campbell, Coker, Neto, and Oezyilmaz]{rosti2014}
Rostislav~A. Doganov, Eoin~C.T. OÕFarrell, Steven~P. Koenig, Yuting Yeo, Kenji
  Watanabe, Takashi Taniguchi, A.~Ziletti, A.~Carvalho, D.K. Campbell, D.~F.
  Coker, A.~H.~Castro Neto, and Barbaros Oezyilmaz.
\newblock \emph{unpublished}.

\bibitem[Castellanos-Gomez et~al.()Castellanos-Gomez, Vicarelli, Prada, Island,
  Narasimha-Acharya, Blanter, Groenendijk, Buscema, Steele, Alvarez,
  Zandbergen, Palacios, and van~der Zant]{castellanos-gomez2014}
Andres Castellanos-Gomez, Leonardo Vicarelli, Elsa Prada, Joshua~O Island, K~L
  Narasimha-Acharya, Sofya~I Blanter, Dirk~J Groenendijk, Michele Buscema,
  Gary~A Steele, J~V Alvarez, Henny~W Zandbergen, J~J Palacios, and Herre S~J
  van~der Zant.
\newblock Isolation and characterization of few-layer black phosphorus.
\newblock \emph{2D Materials}, 1\penalty0 (2):\penalty0 025001.

\bibitem[Favron et~al.()Favron, Gaufres, Fossard, L'evesque, Phaneuf-L'Heureux,
  N.~Y-W.~Tang, Leonelli, Francoeur, and Martel]{favron}
A.~Favron, E.~Gaufres, F.~Fossard, P.L. L'evesque, A-L. Phaneuf-L'Heureux,
  A.~Loiseau N.~Y-W.~Tang, R.~Leonelli, S.~Francoeur, and R.~Martel.
\newblock \emph{arXiv:1408.0345}.

\bibitem[Yau et~al.(1992)Yau, Moffat, Zhang, and Lerner]{stm}
Shueh-Lin Yau, Thomas~P. Moffat, Allen J. Bard~Zhengwei Zhang, and Michael~M.
  Lerner.
\newblock \emph{Chem. Phys. Lett.}, 198:\penalty0 383, 1992.

\bibitem[Farnsworth et~al.(2014)Farnsworth, Wells, Woomer, Hu, Donley, and
  Warren]{farnsworth2014}
Tyler~W Farnsworth, Rebekah~A Wells, Adam~H Woomer, Jun Hu, Carrie Donley, and
  Scott~C Warren.
\newblock Understanding the surface reactivity of 2-d black phosphorus.
\newblock \emph{Informal Phosphore Symposium}, 2014.

\bibitem[Brow(2000)]{brow2000}
Richard~K Brow.
\newblock Review: the structure of simple phosphate glasses.
\newblock \emph{Journal of Non-Crystalline Solids}, 263-264\penalty0
  (0):\penalty0 1 -- 28, 2000.
\newblock ISSN 0022-3093.
\newblock \doi{http://dx.doi.org/10.1016/S0022-3093(99)00620-1}.

\bibitem[Jansen(1986)]{jansen1986}
B.~Jansen, M.;~Luer.
\newblock \emph{Z. Kristallogr.}, 177\penalty0 (0):\penalty0 149 -- 151, 1986.

\bibitem[Arbib et~al.(1996)Arbib, Elouadi, Chaminade, and Darriet]{arbib1996}
El~Hassan Arbib, Brahim Elouadi, Jean~Pierre Chaminade, and Jacques Darriet.
\newblock \{BRIEF\} communication: New refinement of the crystal structure of
  o-p2o5.
\newblock \emph{Journal of Solid State Chemistry}, 127\penalty0 (2):\penalty0
  350 -- 353, 1996.
\newblock ISSN 0022-4596.
\newblock \doi{http://dx.doi.org/10.1006/jssc.1996.0393}.
\newblock URL
  \url{http://www.sciencedirect.com/science/article/pii/S002245969690393X}.

\bibitem[Stachel et~al.(1995)Stachel, Svoboda, and Fuess]{stachel1995}
D.~Stachel, I.~Svoboda, and H.~Fuess.
\newblock {Phosphorus Pentoxide at 233 K}.
\newblock \emph{Acta Crystallographica Section C}, 51\penalty0 (6):\penalty0
  1049--1050, Jun 1995.
\newblock \doi{10.1107/S0108270194012126}.
\newblock URL \url{http://dx.doi.org/10.1107/S0108270194012126}.

\bibitem[Hulliger(1976)]{hullinger1976}
F.~Hulliger.
\newblock \emph{Structural Chemistry of Layer-Type Phases}.
\newblock Springer, 1976.

\bibitem[Galeener and Jr.(1979)]{galeener1979}
F.L. Galeener and J.C.~Mikkelsen Jr.
\newblock The raman spectra and structure of pure vitreous \{P2O5\}.
\newblock \emph{Solid State Communications}, 30\penalty0 (8):\penalty0 505 --
  510, 1979.
\newblock ISSN 0038-1098.
\newblock \doi{http://dx.doi.org/10.1016/0038-1098(79)91227-4}.
\newblock URL
  \url{http://www.sciencedirect.com/science/article/pii/0038109879912274}.

\bibitem[Wang et~al.()Wang, Pandey, Zhang, and Lerner]{wang}
Gaoxue Wang, Ravindra Pandey, Shashi P. Karna~Zhengwei Zhang, and Michael~M.
  Lerner.
\newblock \emph{arXiv:1409.0459}.

\bibitem[Giannozzi et~al.(2009)Giannozzi, Baroni, Bonini, Calandra, Car,
  Cavazzoni, Ceresoli, Chiarotti, Cococcioni, Dabo, {Dal Corso}, de~Gironcoli,
  Fabris, Fratesi, Gebauer, Gerstmann, Gougoussis, Kokalj, Lazzeri,
  Martin-Samos, Marzari, Mauri, Mazzarello, Paolini, Pasquarello, Paulatto,
  Sbraccia, Scandolo, Sclauzero, Seitsonen, Smogunov, Umari, and
  Wentzcovitch]{qe}
Paolo Giannozzi, Stefano Baroni, Nicola Bonini, Matteo Calandra, Roberto Car,
  Carlo Cavazzoni, Davide Ceresoli, Guido~L Chiarotti, Matteo Cococcioni,
  Ismaila Dabo, Andrea {Dal Corso}, Stefano de~Gironcoli, Stefano Fabris, Guido
  Fratesi, Ralph Gebauer, Uwe Gerstmann, Christos Gougoussis, Anton Kokalj,
  Michele Lazzeri, Layla Martin-Samos, Nicola Marzari, Francesco Mauri,
  Riccardo Mazzarello, Stefano Paolini, Alfredo Pasquarello, Lorenzo Paulatto,
  Carlo Sbraccia, Sandro Scandolo, Gabriele Sclauzero, Ari~P Seitsonen,
  Alexander Smogunov, Paolo Umari, and Renata~M Wentzcovitch.
\newblock Quantum espresso: a modular and open-source software project for
  quantum simulations of materials.
\newblock \emph{Journal of Physics: Condensed Matter}, 21\penalty0
  (39):\penalty0 395502 (19pp), 2009.
\newblock URL \url{http://www.quantum-espresso.org}.

\bibitem[Perdew et~al.(1996)Perdew, Burke, and Ernzerhof]{pbe}
John~P. Perdew, Kieron Burke, and Matthias Ernzerhof.
\newblock Generalized gradient approximation made simple.
\newblock \emph{Phys. Rev. Lett.}, 77:\penalty0 3865--3868, Oct 1996.
\newblock \doi{10.1103/PhysRevLett.77.3865}.
\newblock URL \url{http://link.aps.org/doi/10.1103/PhysRevLett.77.3865}.

\bibitem[Perdew et~al.(2008)Perdew, Ruzsinszky, Csonka, Vydrov, Scuseria,
  Constantin, Zhou, and Burke]{PBEsol}
John~P. Perdew, Adrienn Ruzsinszky, G\'abor~I. Csonka, Oleg~A. Vydrov,
  Gustavo~E. Scuseria, Lucian~A. Constantin, Xiaolan Zhou, and Kieron Burke.
\newblock Restoring the density-gradient expansion for exchange in solids and
  surfaces.
\newblock \emph{Phys. Rev. Lett.}, 100:\penalty0 136406, Apr 2008.
\newblock \doi{10.1103/PhysRevLett.100.136406}.
\newblock URL \url{http://link.aps.org/doi/10.1103/PhysRevLett.100.136406}.

\bibitem[Krukau et~al.(2006)Krukau, Vydrov, Izmaylov, and Scuseria]{hse06}
Aliaksandr~V. Krukau, Oleg~A. Vydrov, Artur~F. Izmaylov, and Gustavo~E.
  Scuseria.
\newblock Influence of the exchange screening parameter on the performance of
  screened hybrid functionals.
\newblock \emph{The Journal of Chemical Physics}, 125\penalty0 (22):\penalty0
  224106, 2006.
\newblock \doi{http://dx.doi.org/10.1063/1.2404663}.
\newblock URL
  \url{http://scitation.aip.org/content/aip/journal/jcp/125/22/10.1063/1.2404663}.

\bibitem[Heyd et~al.(2005)Heyd, Peralta, Scuseria, and Martin]{heyd2005}
Jochen Heyd, Juan~E. Peralta, Gustavo~E. Scuseria, and Richard~L. Martin.
\newblock Energy band gaps and lattice parameters evaluated with the
  heyd-scuseria-ernzerhof screened hybrid functional.
\newblock \emph{The Journal of Chemical Physics}, 123\penalty0 (17):\penalty0
  174101, 2005.
\newblock \doi{http://dx.doi.org/10.1063/1.2085170}.
\newblock URL
  \url{http://scitation.aip.org/content/aip/journal/jcp/123/17/10.1063/1.2085170}.

\bibitem[Janesko et~al.(2009)Janesko, Henderson, and Scuseria]{janesko2009}
Benjamin~G. Janesko, Thomas~M. Henderson, and Gustavo~E. Scuseria.
\newblock Screened hybrid density functionals for solid-state chemistry and
  physics.
\newblock \emph{Phys. Chem. Chem. Phys.}, 11:\penalty0 443--454, 2009.
\newblock \doi{10.1039/B812838C}.
\newblock URL \url{http://dx.doi.org/10.1039/B812838C}.

\bibitem[Henderson et~al.(2011)Henderson, Paier, and Scuseria]{henderson2011}
Thomas~M. Henderson, Joachim Paier, and Gustavo~E. Scuseria.
\newblock Accurate treatment of solids with the hse screened hybrid.
\newblock \emph{physica status solidi (b)}, 248\penalty0 (4):\penalty0
  767--774, 2011.
\newblock ISSN 1521-3951.
\newblock \doi{10.1002/pssb.201046303}.
\newblock URL \url{http://dx.doi.org/10.1002/pssb.201046303}.

\bibitem[Bl\"ochl(1994)]{paw}
P.~E. Bl\"ochl.
\newblock Projector augmented-wave method.
\newblock \emph{Phys. Rev. B}, 50:\penalty0 17953--17979, Dec 1994.
\newblock \doi{10.1103/PhysRevB.50.17953}.
\newblock URL \url{http://link.aps.org/doi/10.1103/PhysRevB.50.17953}.

\bibitem[Troullier and Martins(1991)]{tm-pseudo}
N.~Troullier and Jos\'e~Luriaas Martins.
\newblock Efficient pseudopotentials for plane-wave calculations.
\newblock \emph{Phys. Rev. B}, 43:\penalty0 1993--2006, Jan 1991.
\newblock \doi{10.1103/PhysRevB.43.1993}.
\newblock URL \url{http://link.aps.org/doi/10.1103/PhysRevB.43.1993}.

\bibitem[Baroni et~al.(2001)Baroni, de~Gironcoli, Dal~Corso, and
  Giannozzi]{baroni2001}
Stefano Baroni, Stefano de~Gironcoli, Andrea Dal~Corso, and Paolo Giannozzi.
\newblock Phonons and related crystal properties from density-functional
  perturbation theory.
\newblock \emph{Rev. Mod. Phys.}, 73:\penalty0 515--562, Jul 2001.
\newblock \doi{10.1103/RevModPhys.73.515}.
\newblock URL \url{http://link.aps.org/doi/10.1103/RevModPhys.73.515}.

\bibitem[Monkhorst and Pack(1976)]{mpgrid}
Hendrik~J. Monkhorst and James~D. Pack.
\newblock Special points for brillouin-zone integrations.
\newblock \emph{Phys. Rev. B}, 13:\penalty0 5188--5192, Jun 1976.
\newblock \doi{10.1103/PhysRevB.13.5188}.
\newblock URL \url{http://link.aps.org/doi/10.1103/PhysRevB.13.5188}.

\bibitem[Bl\"ochl et~al.(1994)Bl\"ochl, Jepsen, and Andersen]{smear-dos}
Peter~E. Bl\"ochl, O.~Jepsen, and O.~K. Andersen.
\newblock Improved tetrahedron method for brillouin-zone integrations.
\newblock \emph{Phys. Rev. B}, 49:\penalty0 16223--16233, Jun 1994.
\newblock \doi{10.1103/PhysRevB.49.16223}.
\newblock URL \url{http://link.aps.org/doi/10.1103/PhysRevB.49.16223}.

\bibitem[Perdew et~al.(1982)Perdew, Parr, Levy, and Balduz]{perdew1982}
John~P. Perdew, Robert~G. Parr, Mel Levy, and Jose~L. Balduz.
\newblock Density-functional theory for fractional particle number: Derivative
  discontinuities of the energy.
\newblock \emph{Phys. Rev. Lett.}, 49:\penalty0 1691--1694, Dec 1982.
\newblock \doi{10.1103/PhysRevLett.49.1691}.
\newblock URL \url{http://link.aps.org/doi/10.1103/PhysRevLett.49.1691}.

\bibitem[Perdew and Levy(1983)]{perdew1983}
John~P. Perdew and Mel Levy.
\newblock Physical content of the exact kohn-sham orbital energies: Band gaps
  and derivative discontinuities.
\newblock \emph{Phys. Rev. Lett.}, 51:\penalty0 1884--1887, Nov 1983.
\newblock \doi{10.1103/PhysRevLett.51.1884}.
\newblock URL \url{http://link.aps.org/doi/10.1103/PhysRevLett.51.1884}.

\bibitem[Hedin(1965)]{hedin1965}
Lars Hedin.
\newblock New method for calculating the one-particle green's function with
  application to the electron-gas problem.
\newblock \emph{Phys. Rev.}, 139:\penalty0 A796--A823, Aug 1965.
\newblock \doi{10.1103/PhysRev.139.A796}.
\newblock URL \url{http://link.aps.org/doi/10.1103/PhysRev.139.A796}.

\bibitem[Hybertsen and Louie(1986)]{hybertsen1986}
Mark~S. Hybertsen and Steven~G. Louie.
\newblock Electron correlation in semiconductors and insulators: Band gaps and
  quasiparticle energies.
\newblock \emph{Phys. Rev. B}, 34:\penalty0 5390--5413, Oct 1986.
\newblock \doi{10.1103/PhysRevB.34.5390}.
\newblock URL \url{http://link.aps.org/doi/10.1103/PhysRevB.34.5390}.

\bibitem[LOUIE()]{louie1998}
STEVEN~G. LOUIE.
\newblock \emph{FIRST-PRINCIPLES THEORY OF ELECTRON EXCITATION ENERGIES IN
  SOLIDS, SURFACES, AND DEFECTS}, chapter~3, pages 96--142.
\newblock \doi{10.1142/9789812817006_0003}.
\newblock URL
  \url{http://www.worldscientific.com/doi/abs/10.1142/9789812817006_0003}.

\bibitem[Malone and Cohen(2013)]{malone2013}
Brad~D Malone and Marvin~L Cohen.
\newblock Quasiparticle semiconductor band structures including spin-orbit
  interactions.
\newblock \emph{Journal of Physics: Condensed Matter}, 25\penalty0
  (10):\penalty0 105503, 2013.
\newblock URL \url{http://stacks.iop.org/0953-8984/25/i=10/a=105503}.

\bibitem[Ismail-Beigi(2006)]{beigi2006}
Sohrab Ismail-Beigi.
\newblock Truncation of periodic image interactions for confined systems.
\newblock \emph{Phys. Rev. B}, 73:\penalty0 233103, Jun 2006.
\newblock \doi{10.1103/PhysRevB.73.233103}.
\newblock URL \url{http://link.aps.org/doi/10.1103/PhysRevB.73.233103}.

\bibitem[Gonze et~al.(2009)Gonze, Amadon, Anglade, Beuken, Bottin, Boulanger,
  Bruneval, Caliste, Caracas, Cote, Deutsch, Genovese, Ghosez, Giantomassi,
  Goedecker, Hamann, Hermet, Jollet, Jomard, Leroux, Mancini, Mazevet,
  Oliveira, Onida, Pouillon, Rangel, Rignanese, Sangalli, Shaltaf, Torrent,
  Verstraete, Zerah, and Zwanziger]{abinit2009}
X.~Gonze, B.~Amadon, P.M. Anglade, J.M. Beuken, F.~Bottin, P.~Boulanger,
  F.~Bruneval, D.~Caliste, R.~Caracas, M.~Cote, T.~Deutsch, L.~Genovese, Ph.
  Ghosez, M.~Giantomassi, S.~Goedecker, D.R. Hamann, P.~Hermet, F.~Jollet,
  G.~Jomard, S.~Leroux, M.~Mancini, S.~Mazevet, M.J.T. Oliveira, G.~Onida,
  Y.~Pouillon, T.~Rangel, G.M. Rignanese, D.~Sangalli, R.~Shaltaf, M.~Torrent,
  M.J. Verstraete, G.~Zerah, and J.W. Zwanziger.
\newblock Abinit: First-principles approach to material and nanosystem
  properties.
\newblock \emph{Computer Physics Communications}, 180\penalty0 (12):\penalty0
  2582 -- 2615, 2009.
\newblock ISSN 0010-4655.
\newblock \doi{http://dx.doi.org/10.1016/j.cpc.2009.07.007}.
\newblock URL
  \url{http://www.sciencedirect.com/science/article/pii/S0010465509002276}.
\newblock 40 \{YEARS\} \{OF\} CPC: A celebratory issue focused on quality
  software for high performance, grid and novel computing architectures.

\bibitem[Greenwood and Earnshaw(1997)]{greenwood1997}
N~N. Greenwood and A~Earnshaw.
\newblock \emph{Chemistry of the Elements}.
\newblock Reed Elsevier, 1997.

\bibitem[Jansen et~al.(1981)Jansen, Voss, and Deiseroth]{jansen1981}
Martin Jansen, Marlen Voss, and Hans-Jorg Deiseroth.
\newblock Structural properties of phosphorus oxides in the solid aggregation
  state.
\newblock \emph{Angewandte Chemie International Edition in English},
  20\penalty0 (11):\penalty0 965--965, 1981.
\newblock ISSN 1521-3773.
\newblock \doi{10.1002/anie.198109651}.
\newblock URL \url{http://dx.doi.org/10.1002/anie.198109651}.

\bibitem[Walker et~al.(1979)Walker, Peckenpaugh, and Mills]{walker1979}
Michael~L. Walker, Daniel~E. Peckenpaugh, and Jerry~L. Mills.
\newblock Mixed phosphorus(iii)-phosphorus(v) oxide chalcogenides.
\newblock \emph{Inorganic Chemistry}, 18\penalty0 (10):\penalty0 2792--2796,
  1979.
\newblock \doi{10.1021/ic50200a032}.
\newblock URL \url{http://pubs.acs.org/doi/abs/10.1021/ic50200a032}.

\bibitem[Clade et~al.(1994)Clade, Frick, and Jansen]{clade1994}
J.~Clade, F.~Frick, and M.~Jansen.
\newblock Recent synthetic, structural, spectroscopic, and theoretical studies
  on molecular phosphorus oxides and oxide sulfides.
\newblock 41:\penalty0 327 -- 388, 1994.
\newblock ISSN 0898-8838.
\newblock \doi{http://dx.doi.org/10.1016/S0898-8838(08)60175-0}.
\newblock URL
  \url{http://www.sciencedirect.com/science/article/pii/S0898883808601750}.

\bibitem[Gillespie and Nyholm(1957)]{gillespie1957}
R.~J. Gillespie and R.~S. Nyholm.
\newblock Inorganic stereochemistry.
\newblock \emph{Q. Rev. Chem. Soc.}, 11:\penalty0 339--380, 1957.
\newblock \doi{10.1039/QR9571100339}.
\newblock URL \url{http://dx.doi.org/10.1039/QR9571100339}.

\bibitem[Gillespie(1963)]{gillespie1963}
R.~J. Gillespie.
\newblock The valence-shell electron-pair repulsion (vsepr) theory of directed
  valency.
\newblock \emph{Journal of Chemical Education}, 40\penalty0 (6):\penalty0 295,
  1963.
\newblock \doi{10.1021/ed040p295}.
\newblock URL \url{http://pubs.acs.org/doi/abs/10.1021/ed040p295}.

\bibitem[Gillespie(1992)]{gillespie1992}
Ronald~J. Gillespie.
\newblock The vsepr model revisited.
\newblock \emph{Chem. Soc. Rev.}, 21:\penalty0 59--69, 1992.
\newblock \doi{10.1039/CS9922100059}.
\newblock URL \url{http://dx.doi.org/10.1039/CS9922100059}.

\bibitem[Daia and Zenga()]{daia2014}
Jun Daia and Xiao~Cheng Zenga.
\newblock Structure and stability of two dimensional phosphorene with =o or =nh
  functionalization.
\newblock \emph{arXiv:1409.7719}.

\bibitem[Rudenko and Katsnelson(2014)]{rudenko2014}
A.~N. Rudenko and M.~I. Katsnelson.
\newblock Quasiparticle band structure and tight-binding model for single- and
  bilayer black phosphorus.
\newblock \emph{Phys. Rev. B}, 89:\penalty0 201408, May 2014.
\newblock \doi{10.1103/PhysRevB.89.201408}.
\newblock URL \url{http://link.aps.org/doi/10.1103/PhysRevB.89.201408}.

\bibitem[Valentim et~al.(1997)Valentim, Engels, Peyerimhoff, Clade, and
  Jansen]{valentim1997}
A.~R.~S. Valentim, B.~Engels, S.~Peyerimhoff, J.~Clade, and M.~Jansen.
\newblock Study of the p4o7, p4o6s, and p4o6se vibrational spectra.
\newblock \emph{Inorganic Chemistry}, 36\penalty0 (11):\penalty0 2451--2457,
  1997.
\newblock \doi{10.1021/ic961202s}.
\newblock PMID: 11669885.

\bibitem[Valentim et~al.(1998)Valentim, Engels, Peyerimhoff, Clade, and
  Jansen]{valentim1998}
A.~R.~S. Valentim, B.~Engels, S.~D. Peyerimhoff, J.~Clade, and M.~Jansen.
\newblock A comparative study of the bonding character in the p4on (n = 6?10)
  series by means of a vibrational analysis.
\newblock \emph{The Journal of Physical Chemistry A}, 102\penalty0
  (21):\penalty0 3690--3696, 1998.
\newblock \doi{10.1021/jp9805611}.

\bibitem[Mielke and Andrews(1989)]{mielke1989}
Zofia Mielke and Lester Andrews.
\newblock Infrared spectra of phosphorus oxides (p4o6, p4o7, p4o8, p4o9 and
  p4o10) in solid argon.
\newblock \emph{The Journal of Physical Chemistry}, 93\penalty0 (8):\penalty0
  2971--2976, 1989.
\newblock \doi{10.1021/j100345a024}.

\bibitem[Sugai and Shirotani(1985)]{sugai1985}
S.~Sugai and I.~Shirotani.
\newblock Raman and infrared reflection spectroscopy in black phosphorus.
\newblock \emph{Solid State Communications}, 53\penalty0 (9):\penalty0 753 --
  755, 1985.
\newblock ISSN 0038-1098.
\newblock \doi{http://dx.doi.org/10.1016/0038-1098(85)90213-3}.
\newblock URL
  \url{http://www.sciencedirect.com/science/article/pii/0038109885902133}.

\bibitem[Galeener et~al.(1978)Galeener, Mikkelsen, Geils, and
  Mosby]{galeener1978}
F.~L. Galeener, J.~C. Mikkelsen, R.~H. Geils, and W.~J. Mosby.
\newblock The relative raman cross sections of vitreous sio2, geo2, b2o3, and
  p2o5.
\newblock \emph{Applied Physics Letters}, 32\penalty0 (1), 1978.

\bibitem[Brazhkin et~al.(2011)Brazhkin, Akola, Katayama, Kohara, Kondrin,
  Lyapin, Lyapin, Tricot, and Yagafarov]{brazhkin2011}
V.~V. Brazhkin, J.~Akola, Y.~Katayama, S.~Kohara, M.~V. Kondrin, A.~G. Lyapin,
  S.~G. Lyapin, G.~Tricot, and O.~F. Yagafarov.
\newblock Densified low-hygroscopic form of p2o5 glass.
\newblock \emph{J. Mater. Chem.}, 21:\penalty0 10442--10447, 2011.
\newblock \doi{10.1039/C1JM10889A}.
\newblock URL \url{http://dx.doi.org/10.1039/C1JM10889A}.

\bibitem[Gilheany(1994)]{gilheany1994}
Declan~G. Gilheany.
\newblock Ylides, phosphoniumno d orbitals but walsh diagrams and maybe banana
  bonds: Chemical bonding in phosphines, phosphine oxides, and phosphonium
  ylides.
\newblock \emph{Chemical Reviews}, 94\penalty0 (5):\penalty0 1339--1374, 1994.
\newblock \doi{10.1021/cr00029a008}.
\newblock URL \url{http://pubs.acs.org/doi/abs/10.1021/cr00029a008}.

\bibitem[Rai and Symons(1994)]{rai1994}
Uma~S. Rai and Martyn C.~R. Symons.
\newblock Epr data do not support the p[double bond{,} length as m-dash]o
  representation for trialkyl phosphates and phosphine oxides or sulfides.
\newblock \emph{J. Chem. Soc.{,} Faraday Trans.}, 90:\penalty0 2649--2652,
  1994.
\newblock \doi{10.1039/FT9949002649}.
\newblock URL \url{http://dx.doi.org/10.1039/FT9949002649}.

\bibitem[Chesnut and Savin(1999)]{chesnut1999}
D.~B. Chesnut and A.~Savin.
\newblock The electron localization function (elf) description of the po bond
  in phosphine oxide.
\newblock \emph{Journal of the American Chemical Society}, 121\penalty0
  (10):\penalty0 2335--2336, 1999.
\newblock \doi{10.1021/ja984314m}.
\newblock URL \url{http://pubs.acs.org/doi/abs/10.1021/ja984314m}.

\end{thebibliography}
\bibliographystyle{unsrtnat}	

\end{document}